\documentclass[letter]{aa} 

\usepackage{graphicx}
\usepackage{txfonts}
\usepackage{amstext}

\begin{document} 

\title{Resolving the MYSO binaries PDS 27 and PDS 37 with VLTI/PIONIER}

\author{E. Koumpia \inst{1}, K.~M. Ababakr \inst{10}, W.~J. de Wit \inst{2}, R.~D. Oudmaijer 
     \inst{1}, A. Caratti o Garatti \inst{3}, P. Boley \inst{4,5}, H. Linz \inst{6}, S. Kraus \inst{7}, J.~S. Vink \inst{8}, J.-B Le Bouquin \inst{9} \fnmsep\thanks{Based on ESO observations 094.C-0359, 098.C-0636}
          }
   \institute{School of Physics and Astronomy, E.C. Stoner Building, University of Leeds, Leeds LS2 9JT, UK
              \\
              \email{ev.koumpia@gmail.com}
              \and
              ESO Vitacura, Alonso de C{\'o}rdova 3107 Vitacura, Casilla 19001 Santiago de Chile, Chile 
              \and
              Dublin Institute for Advanced Studies, Astronomy \& Astrophysics Section, 31 Fitzwilliam Place, Dublin 2, Ireland
              \and
              Moscow Institute of Physics and Technology, 9 Institutski per., Dolgoprudny 141701, Russia
              \and
              Ural Federal University, 51 Lenin Ave., Ekaterinburg 620075, Russia
              \and
              Max Planck Institute for Astronomy, K{\"o}nigstuhl 17, 69117 Heidelberg, Germany
              \and 
              University of Exeter, School of Physics, Astrophysics Group, Stocker Road, Exeter, EX4 4QL, UK
              \and 
              Armagh Observatory and Planetarium, College Hill, Armagh BT61 9DG, Ireland
              \and 
              Universit{\'e} Grenoble Alpes, CNRS, IPAG, 38000 Grenoble, France
\and 
              Erbil Polytechnic University, Kirkuk Road, Erbil, Iraq
\\
             }

\date{Received; accepted}

\abstract
{Binarity and multiplicity appear to be a common outcome in star formation. In particular, the binary fraction of massive (OB-type) stars can be very high. In many cases, the further stellar evolution of these stars is affected by binary interactions at some stage during their lifetime. The origin of this high binarity and the binary parameters are poorly understood because observational constraints are scarce, which is predominantly due to a dearth of known young massive binary systems.}{We aim to identify and describe massive young binary systems in order to fill in the gaps of our knowledge of primordial binarity of massive stars, which is crucial for our understanding of massive star formation.}{We observed the two massive young stellar objects (MYSOs) PDS 27 and PDS 37 at the highest spatial resolution provided by VLTI/PIONIER in the H-band (1.3 mas). We applied geometrical models to fit the observed squared visibilities and closure phases. In addition, we performed a radial velocity analysis using published VLT/FORS2 spectropolarimetric and VLT/X-shooter spectroscopic observations.}{Our findings suggest binary companions for both objects at 12 mas (30 au) for PDS 27 and at 22-28 mas (42-54 au) for PDS 37. This means that they are among the closest MYSO binaries resolved to date.}{Our data spatially resolve PDS 27 and PDS 37 for the first time, revealing two of the closest and most massive ($>$8 M$_\odot$) YSO binary candidates to date. PDS 27 and PDS 37 are rare but great laboratories to quantitatively inform and test the theories on formation of such systems.}

\keywords{techniques: interferometric, stars: individual: PDS 27, stars: individual: PDS 37, stars: pre-main-sequence, binaries: close, stars: formation}

\titlerunning{Resolving the MYSO binaries PDS 27 and PDS 37 with PIONIER} 
\authorrunning{Koumpia et al.}
   \maketitle
%

\section{Introduction}

In recent years, substantial progress has been made in the field of
high-mass star formation. The various theoretical scenarios proposed for high-mass star formation suggest that mass is accreted through a circumstellar disk \citep{Jijina1996, Yorke2002, Krumholz2009, kuiper10}. Observational evidence for circumstellar disks has been accumulating using a variety of methods, including AMBER interferometry \citep{Kraus2010}, spectroscopy in the IR wavelength regime \citep{Wheelwright2010,Ilee2013,Ilee2018}, and spectropolarimetry \citep{Ababakr2015}. Submillimeter (submm) interferometric observations indicate the presence of Keplerian disks \citep{Ilee2016,Johnston2015}, while hydrodynamical simulations establish that disks are viable agents for accretion \citep[e.g.,][]{Rosen2016,Klassen2016,Kuiper2018}. For a review on the topic, see \citet{Beltran2016,Hartmann2016}.

Although the formation of massive individual stars is becoming more and more established in both theory and observations, studies have shown that the majority of stars do not form in isolation. In particular, binarity and higher-order multiplicity are fairly well established as a common outcome of star formation. Multiple systems with separations as large as several hundred astronomical units (au) are mostly predicted by numerical simulations as a result of fragmentation processes during the collapse phase \citep[e.g.,][]{Krumholz2012,Myers2013}, while closer binaries ($<$ 10 au) may rather form through accretion disk fragmentation \citep{Meyer2018} or orbital decay during internal \citep[e.g., capture in competitive accretion, magnetic braking during accretion;][]{Bonnell2005,Lund2018} or external (e.g., with other stars) interactions \citep[e.g.,][]{Bate2002}. However, despite these theoretical findings, we should note that reproducible quantitative predictions of binary properties are currently lacking. Observational studies of binaries and multiple systems in the pre-main sequence (PMS) phase are crucial to verify the several theories of their formation \citep[e.g.,][]{Moe2017}.

It has become clear that the multiplicity and binarity of OB-type populations is very high, with a binary fraction between 50-100\% \citep[see also][GRAVITY collaboration]{Chini2012,Sana2013,Sana2014,Karl2018}. Although the evolution and fate of high-mass stars is rather complex, previous studies have shown that they are influenced to a large degree by their binary properties \citep{Sana2012}. Consequently, the multiplicity of high-mass stars is key to understanding their formation and evolution. Our understanding on the high multiplicity rate and the quantitative binary properties during the formation phases is still far from complete however. 

To test the several formation mechanisms of massive young stellar object (MYSO) binaries, it is crucial to be able to detect them during their formation. To date, only a few studies have been dedicated to the multiplicity of MYSOs. The largest survey investigations that come closest in mass and evolutionary phase are reported in \citet{Baines2006}, who studied the intermediate-mass Herbig Ae/Be stars and found a binary fraction of 70\% for separations 0.1 - 1.5 arcsec (50 - 750 au at 500 pc), with an indication that the more massive Herbig Ae/Be stars have a slightly higher binary fraction \citep[see also][]{Pomohaci2019}. Later, \citet{Wheelwright2010} probed the mass ratios of this sample and found rather high mass ratios, suggesting similar masses for both components, whereas \citet{Karl2018} found that the mass ratio distribution declines steeply with mass. \citet{Wheelwright2011} showed that the orbital planes of HAe/Be binaries and the primary disks are likely to be coplanar, favoring the disk fragmentation theories. \citet{Karl2018} reported a high-mass ($>$16~M$_\odot$) binarity fraction of 100\% in a sample of 16 MYSOs, yet previous studies using interferometric observations with the Astronomical Multi-BEam combineR, AMBER, and the Jansky Very Large Array, JVLA, have revealed only a handful of Herbig or MYSO binaries with ranging separations of up to several hundred au and coexistence of disks surrounding the objects \citep[V921 Sco (45 au), NGC7538 IRS2 (700 au), NGC 7538IRS1 (430 au), and IRAS 17216-3801 (170 au),][]{Kraus2012,Kraus2006,Beuther2017,Kraus2017}.

Increasing the number of binary massive YSOs to a healthy statistical value is crucial for our understanding of the formation of massive star systems. Here we focus on two young stellar objects, PDS 27 and PDS 37, which appear to be excellent candidates of massive YSO binaries. \citet{Ababakr2015} presented evidence that they are early-type PMS Herbig Ae/Be stars. Both
objects show a rich emission line spectrum, and they also show a significant near-infrared excess that is due to circumstellar dust. The targets are both hot, 17500~$\pm$ 3500~K, with an inferred spectral type of B2. When combined with
their high luminosities of log(L$_*$/L$_\odot) \sim 4$ \citep{Ababakr2015}, they are placed in a secluded area on the Hertzsprung-Russell (HR) diagram where very young objects lie and are characterized by masses of 12 M$_\odot$ (PDS 27; d=2.55 kpc) and 11 M$_\odot$ (PDS 37; d=1.93 kpc) \citep[GAIA;][]{Vioque2018}. Both stars straddle the intermediate-mass and optically visible Herbig Ae/Be stars class and the more embedded, more massive YSOs. PDS 27 (G231.7986-01.9682) and PDS 37 (G282.2988-00.7769) were included in the RMS survey for MYSOs in the Galaxy, having passed several stringent tests in order to be included in the sample of MYSOs \citep{Lumsden2013}. Given the confusion regarding young and more evolved objects, it is worth mentioning that PDS 37 was included as a post-AGB star in \citet{Szczerba2007}, while \citet{Vieira2011}, based on circumstellar emission, spatial distribution, spectral features and optical/infrared colours suggested that PDS 27 could be a post-AGB object. In this study we rely on the criteria of the \citet{Lumsden2013} survey (e.g., luminosity and color) that consider both objects as YSOs. 


In this paper we present high spatial resolution data obtained with the PIONIER\footnote{Precision Integrated-Optics Near-infrared Imaging ExpeRiment} at the Very Large Telescope (VLTI) in the {\it H} band, and show good evidence that both objects are in fact binary systems. We find close separations (30 au $<$ $\alpha$ $<$ 45 au) at the reported distances, which correspond to the separation of an observed massive YSO binary at the smallest scales that have been spatially resolved so far (together with V921 Sco). In Sec. 2 we present our observations and data reduction. Sec. 3 presents the geometrical models and results, and we conclude with a discussion of the results in Sec. 4.

\vspace{-0.1cm}
\section{Observations}
\label{c3:sec:int}
{\bf{Interferometry}}. The near-infrared interferometric observations of PDS 27 and PDS 37 were obtained with PIONIER \citep{LeBouquin2011}, an H-band (1.66\,$\mu$m) four-beam combiner, at the VLTI using the four 8.2 m Unit Telescopes (UT). The interferometric observables are thus simultaneously measured on six baselines and closure phases for three independent triangles. Two observations were made for PDS 27 during the nights of 2015 March 2 and 3. Only two closure phases are available for the first night of PDS 27. The data of PDS 37 were taken on the night of 2015 March 3. The spectral capabilities were limited to one spectral channel ($\lambda$/$\Delta\lambda$ $\sim$ 5, bandwidth$\sim$0.3~${\rm\mu}$m) across the H band. The projected baseline lengths, B, ranging from $\sim$ 40 to 130 m, correspond to a resolution range of $\sim$ 4.2 to 1.3 mas, respectively, at the wavelength of 1.66~${\rm\mu}$m ($\lambda$/2B in rad). This subsequently corresponds to structures of $\sim$3.3--11~au for PDS 27 and $\sim$2.5--8~au for PDS 37 at the distances of the sources. The uv-plane coverages of the two objects are shown in Figure~\ref{obs_uv}. PIONIER uses single-mode fibers, therefore the field of view corresponds to the point spread function (PSF) of the telescope delivered at the fiber injection point, which is $\sim$50~mas for the UTs in the H band. The seeing during the first observing night varied between 0.78'' and 0.94'' and coherence times ranged between 6 ms and 7.4 ms. For the second observing night, the conditions were slightly worse, with corresponding values of seeing at 0.82-1'' and coherence times of 4.2-5.3 ms. HIP35943 and HIP50026 with H magnitudes of 7 and 7.9 and limb-darkened diameters of 0.25~mas and 0.08~mas, respectively \citep[JMMC SearchCal;][]{Bonneau2011}, were used as standard calibrator objects for PDS 27 and PDS 37, respectively, within a radius of 80$\arcmin$ of the science objects and under relatively stable weather conditions. 

The data were reduced using the automated PIONIER pipeline, which was
developed by Le Bouquin \citep[PNDRS\footnote{\texttt{http://www.jmmc.fr/data\_processing\_pionier.htm}};][]{LeBouquin2011}. Our interferometric observables consist of the squared visibilities (V$^{2}$; power spectrum) and the closure phases (bispectrum). The observables as a function of baseline length at the wavelength of 1.66~${\rm\mu}$m toward PDS~27 and PDS~37, together with the technical overview of the observations, are presented in Table~\ref{observables}. 

\vspace{0.3cm}

{\bf{Spectroscopy}}. We used the spectral information of PDS 27 and PDS 37 provided by the medium-resolution spectrograph X-Shooter on the VLT and FORS2 on the VLT, presented in \citet{Ababakr2015} and in a forthcoming paper by Oudmaijer et al. Interestingly, the Fe II and Ca II emission lines revealed a velocity variation of 18$\pm$8 kms$^{-1}$ in a period of$\text{}$ about two years, while the more recent X-Shooter observations of the object show no velocity offset in a total period of $\text{about seven}$ years compared with the former X-Shooter observations taken in two different epochs. We performed a radial velocity analysis and present estimates at four different epochs in Sec.~\ref{spec}.

\section{Analysis and results}

\subsection{Observational results}

The calibrated observed squared visibility for both objects, even at the smallest baseline, is not equal to one, indicating that either the objects are not point sources but extended, or that they are a member of a binary system at the resolution of our data. In addition, the squared visibility does not systematically decrease with baseline length but fluctuates strongly (see results in Figs.~\ref{models} and~\ref{models_27}). This could suggest that the objects are surrounded by elongated structures (e.g., a flattened ring) and/or that they are binaries (see sec.\ref{geo}). The observed tapering in visibilities may be affected by bandwidth smearing, especially toward the longer baselines, mimicking resolved components (see Appendix~\ref{smear}).

Both objects are characterized by near-zero observed closure phases. In particular, PDS 37 shows large uncertainties in the closure phases, but PDS 27 has a non-zero closure phase at a 2.5$\sigma$ level. For an equal-brightness binary of two point sources, as for any point-symmetric brightness distribution, one expects closure phases of $0^{\circ}$ or $180^{\circ}$. We observed closure phases close to $0^{\circ}$ but not $180^{\circ}$. For strongly unequal brightness binaries, the closure phase in radians is roughly on the same order of magnitude as the flux ratio \citep{Monnier2003}, indicating the amount of the asymmetry. The observed closure phases suggest that the two objects are not entirely equal but are not very high-contrast binaries either, with PDS 27 showing a higher asymmetry in brightness than PDS 37. The closure phases are generally sensitive to the PA of the binary, the separation, and the brightness ratio of the binary components. The situation can be even more complex in case of a resolved binary in which one or both components are also resolved but are characterized by different sizes \citep[see, e.g.,][]{Kraus2017}.

\subsection{Geometrical models}
\label{geo}

To fit the observed visibilities and closure phases of PDS 27 and PDS 37, we used the model fitting software LITpro\footnote{LITpro is developed and maintained by the Jean-Marie Mariotti Center (JMMC) http://www.jmmc.fr/litpro} \citep{Tallon2008}. The parameterization, optimization process including error estimates, and the progression on fitting the visibilities from simple single geometries to binary models for the two objects are described in Appendix~\ref{exploring}. The simpler models (e.g., single symmetric geometries) provide us with a poor fit and very large ${\chi}^2$ (>100). Therefore we proceeded with the more complex geometries described below. 

\begin{figure*}[h]
\begin{center}$ 
\begin{array}{ccc}
  \includegraphics[scale=0.4,trim={50 50 70 0},clip=true]{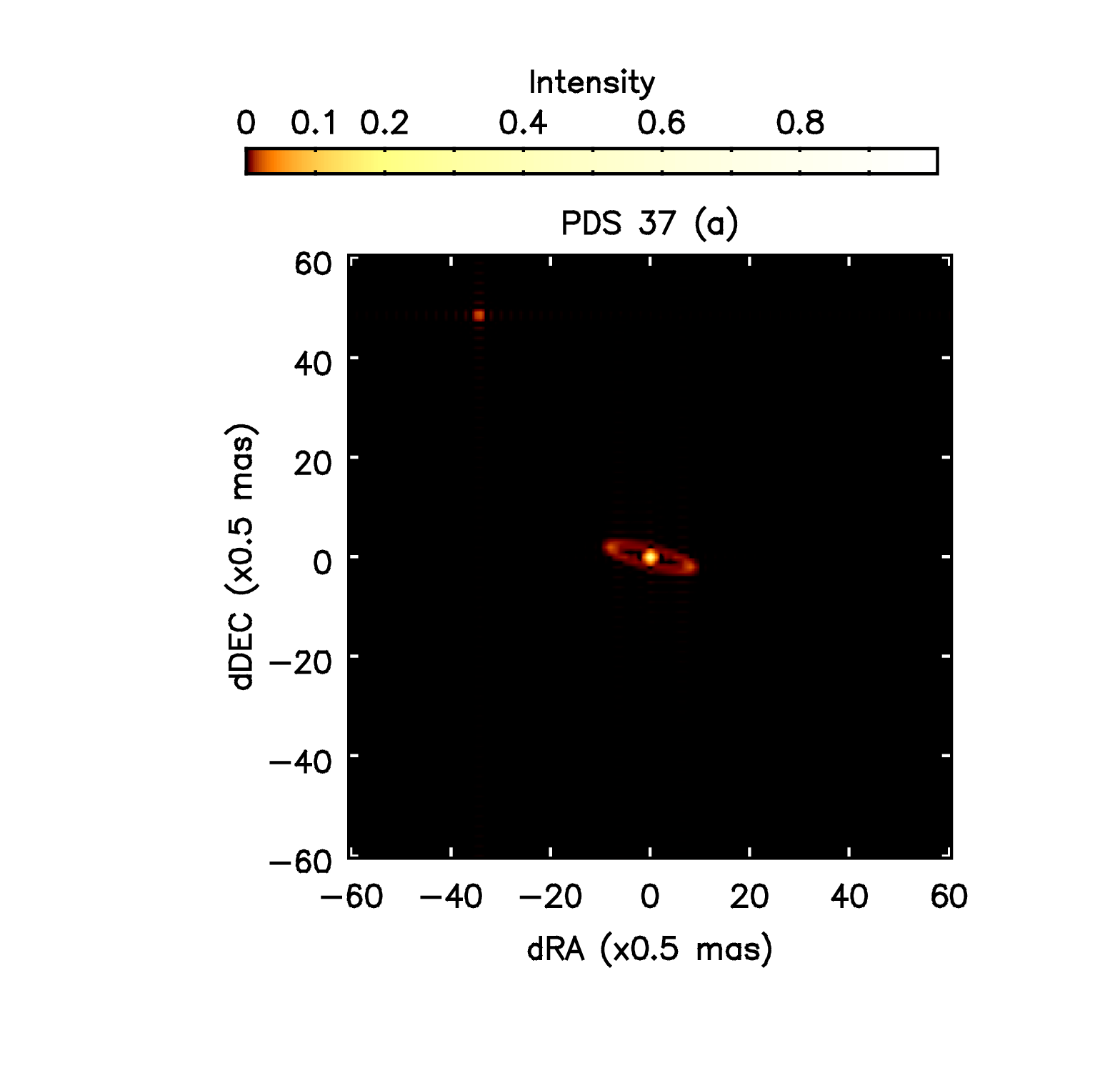} &
  \includegraphics[scale=0.4,trim={50 50 70 0},clip=true]{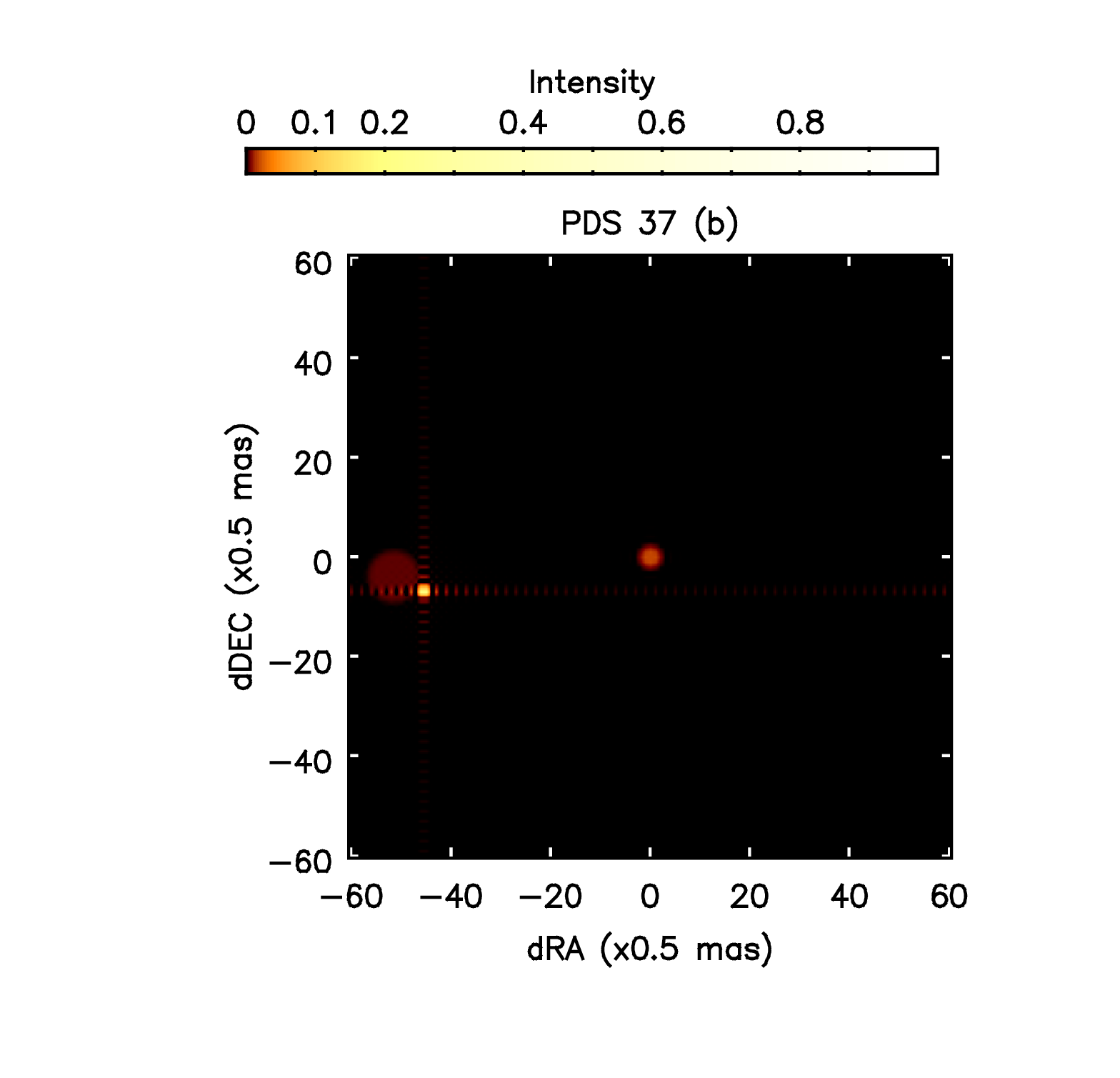} &
  \includegraphics[scale=0.4,trim={50 50 0 0},clip=true]{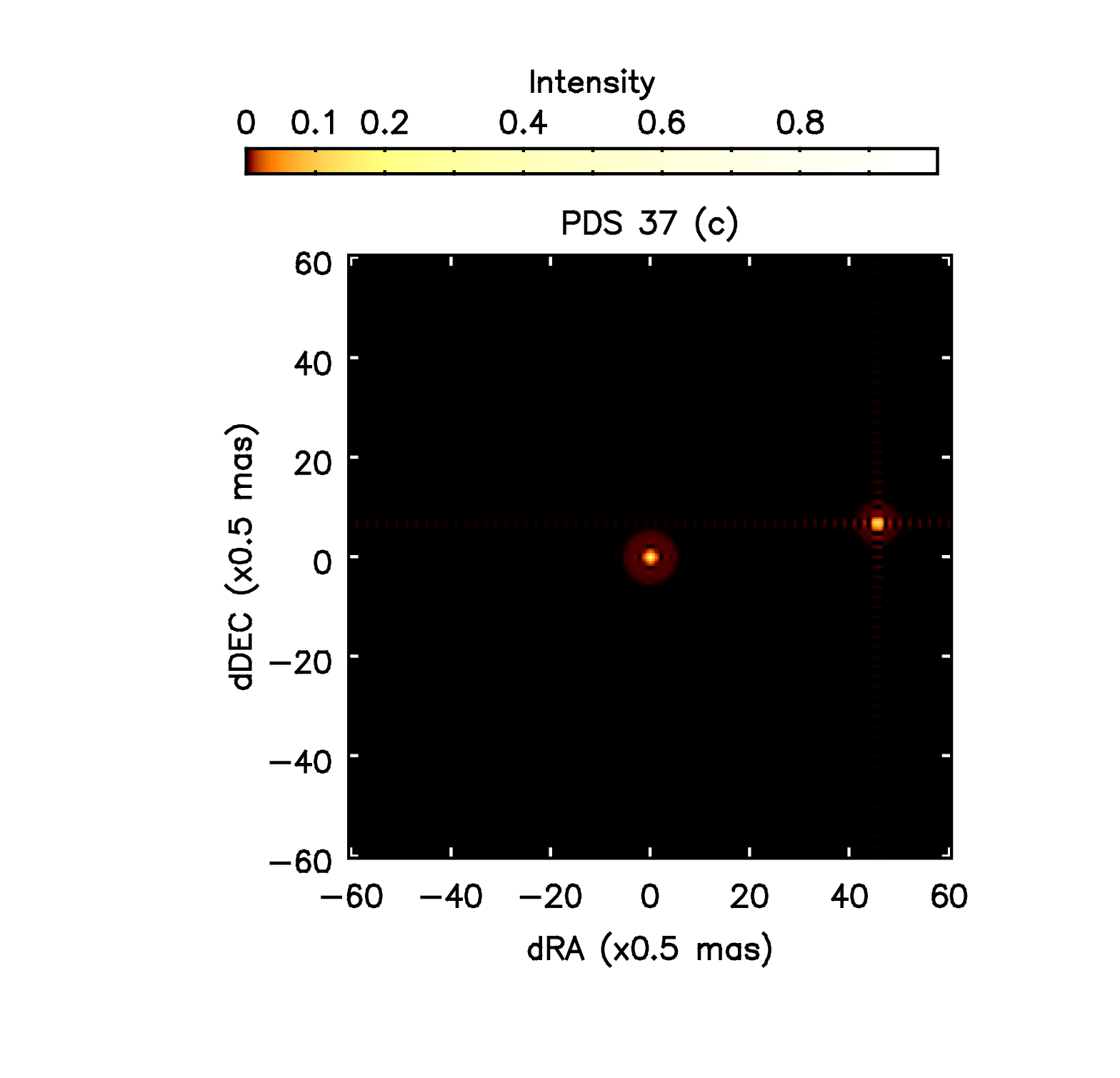} \\
\end{array}$
\end{center}
\begin{center}$
\begin{array}{cc}
\includegraphics[width=6cm,trim={0 0 0 0},clip=true]{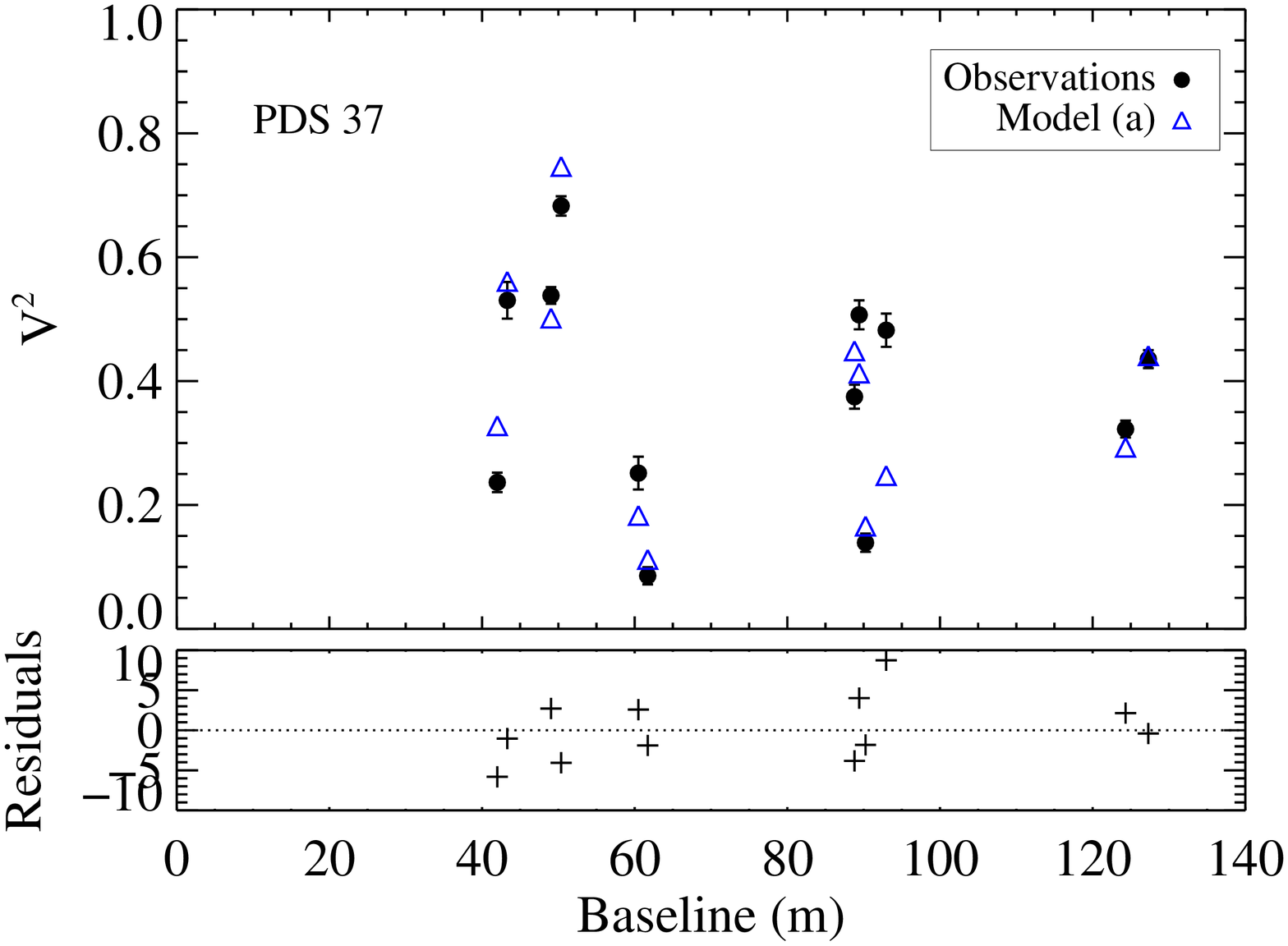} &
 \includegraphics[width=6cm,trim={0 0 0 0},clip=true]{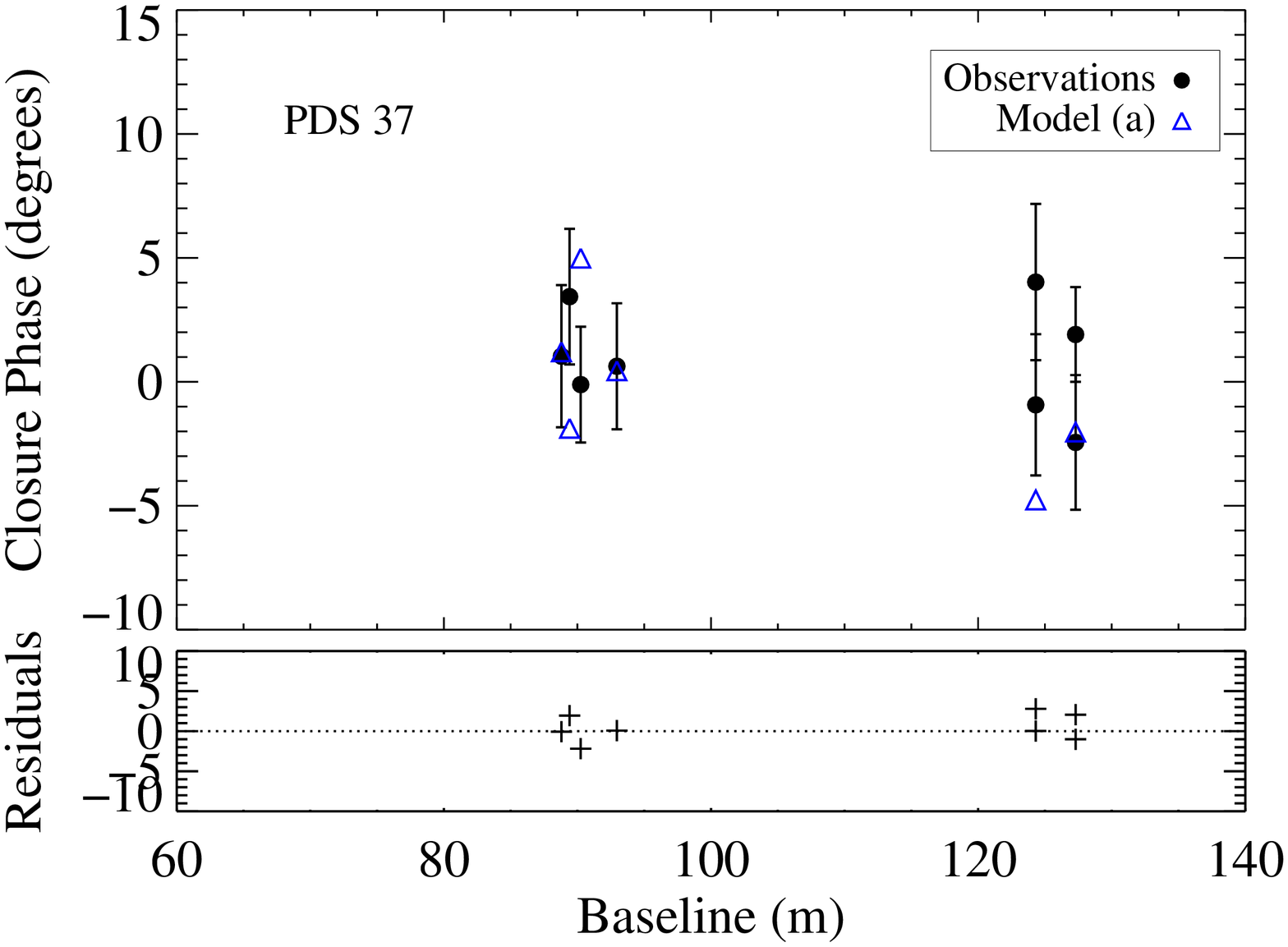} \\
\end{array}$
\end{center}
          \caption{Three different geometrical models are compared with the observed visibilities (V$^{2}$) and closure phases of PDS 37. The black points with vertical error bars are the observed data, and the blue triangles correspond to a representative best-fit model (a). The best-fit models correspond to a) a binary with a ring surrounding the primary (Red. ${\chi}^2$ $\sim$10), b) a binary with both companions resolved and some asymmetric flux toward its primary (Red. ${\chi}^2$ $\sim$6), and c) a binary model with both companions resolved (Red. ${\chi}^2$=10).}
   \label{models}
       \end{figure*}

{\bf{PDS 37.}} We constructed several binary models to fit both visibilities and closure phases (Figure~\ref{models}). The best-fit models are a) a binary with a ring surrounding the primary and an unresolved companion (${\chi}^2$ $\sim$ 10), b) a binary with two resolved components showing asymmetric flux toward its primary (${\chi}^2$ $\sim$ 6), and c) a binary with two resolved components (uniform disks; (${\chi}^2$ $\sim$ 10)). The resulting parameters are presented in Table~\ref{results2}. We find that the separation is reasonably constrained to be $\sim$ 22-28 mas (42-54 au) for PA of 80$^{\circ}$, 260$^{\circ}$ , and 305$^{\circ}$. However, the flux ratios are highly uncertain, and although their values decrease with separation, we cannot make robust conclusions on this effect because the models are not the same (Appendix~\ref{flux_dist_bin_1}). The different values of PA for the different models reflect the uncertainty on the exact position of the secondary object (i.e., presence of more than one antisymmetric minimum).

\begin{table*}[ht]
\caption{Parameters of the models for PDS 37 and PDS 27.}
\small
\centering
\setlength\tabcolsep{2pt}
\begin{tabular}{l l l l l l l l l l l}
\hline\hline
Model & Flux weight 1 & Flux w. 2 & Flux w. ring & Inner diameter & Width & PA (minor axis) & Flatten ratio & x2 & y2 &  \\ ring & & & & mas & mas & degrees & & mas & mas &  \\
\hline
{\bf{PDS 37 (a)}} & 0.5$\pm$0.4 & 0.06$\pm$0.05 & 0.44$\pm$0.34 & 7.7$\pm$0.3 & 0.78$\pm$0.25 & 14.0$\pm$1.4 & 5$\pm$1 & -17.2$\pm$0.3 & 24.3$\pm$0.4 & \\
{\bf{PDS 27 (a)}} & 0.5$\pm$0.4 & & 0.5$\pm$0.4 & 2.6$\pm$0.4 & 4.7$\pm$0.2 & 12.6$\pm$1.2 & 8$\pm$5 & 0 & 0 & \\
\hline
Model & Flux w. disk 1 & Flux w. 1 & Flux w. disk 2 & Flux w. 2 & Diameter 1 & Diameter 2 & x1 & y1 & x2 & y2 \\ disk & & & & & mas & mas & mas & mas & mas & mas \\
\hline
{\bf{PDS 37 (b)}} & 0.2$\pm$0.1 & $0.001^{+0.054}_{-0.001}$ & 0.28$\pm$0.15 & 0.5$\pm$0.3 & 1.9$\pm$0.2 & 5$\pm$0.3 & -25.8$\pm$0.2 & y2=-2.0$\pm$0.3 & -22.7$\pm$0.1 & -3.4$\pm$0.1 \\
{\bf{PDS 27 (b)}} & 0.33$\pm$0.3 & 0.49$\pm$0.44 & 0.18$\pm$0.16 & 0 & 5.4$\pm$0.3 & 3.0$\pm$0.3 & 0 & 0 & -4.0$\pm$0.2 & -11.5$\pm$0.3 \\
{\bf{PDS 37 (c)}} & 0.22$\pm$0.18 & 0.35$\pm$0.28 & 0.11$\pm$0.11 & 0.32$\pm$0.25 & 1.9$\pm$0.2 & 5$\pm$0.3 & 22.8$\pm$0.1 & 3.4$\pm$0.1 & 22.8$\pm$0.1 & 3.4$\pm$0.1 \\
\hline
\label{results2}
\end{tabular}
\tiny {\bf{Notes}}: For PDS 37 the reduced ${\chi}^2$[DoF] (degrees of freedom) for the best-fit models a, b, and c are 10[11], 16[10], and 10[10], respectively, while for PDS 27, the corresponing values for the best-fit models a and b are 10[12] and 12[11], respectively. Both visibilities and closure phases were taken into account.
\end{table*}


Although we cannot clearly distinguish the best solution among our models and remove several ambiguities, we have a good indication that PDS 37 consists of at least two probably resolved components with a separation of 42-54 au.

\begin{figure*}[h]
\begin{center}$ 
\begin{array}{ccc} 
  \includegraphics[scale=0.4,trim={50 -80 70 0},clip=true]{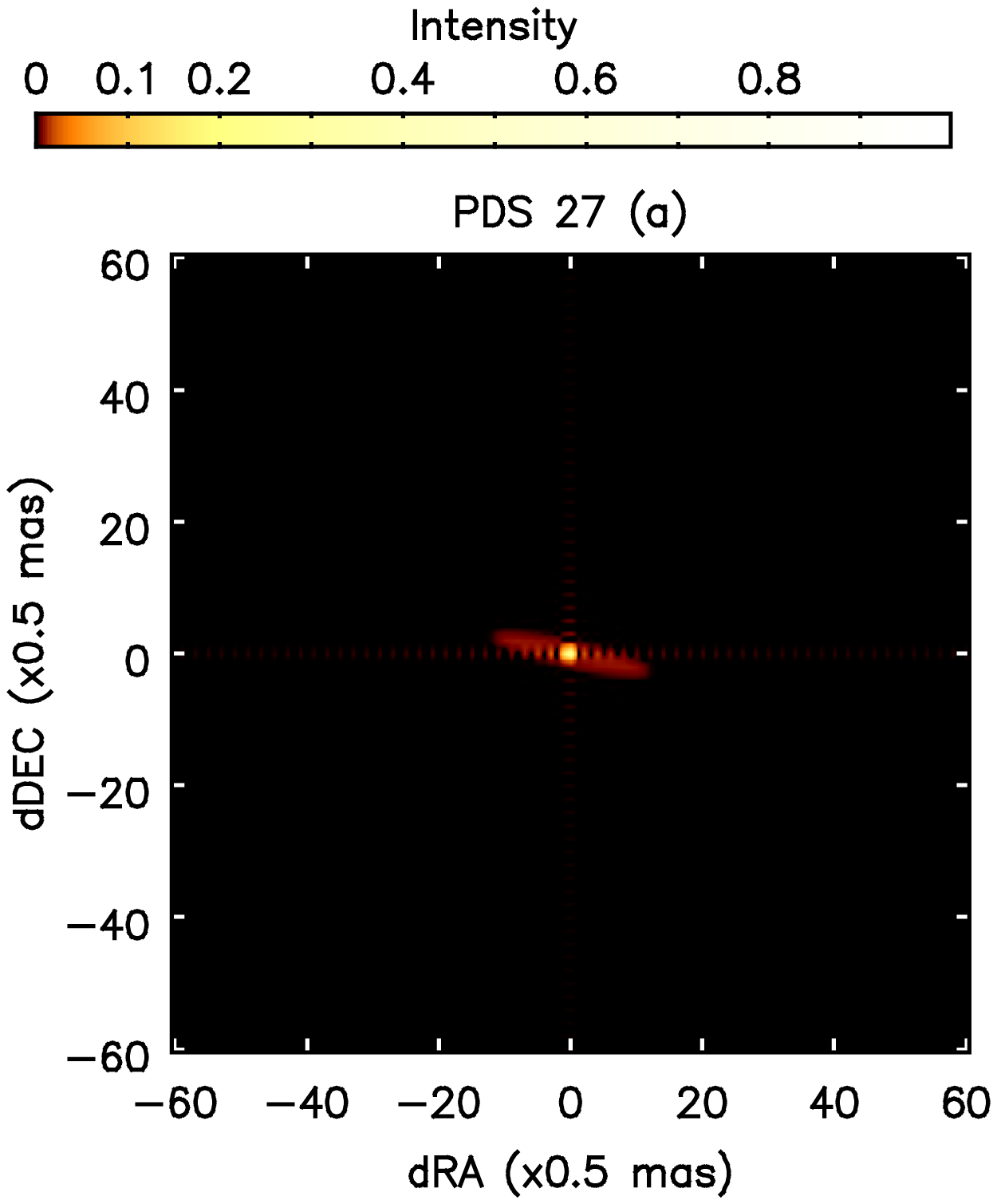} &
  \includegraphics[scale=0.4,trim={50 -80 70 0},clip=true]{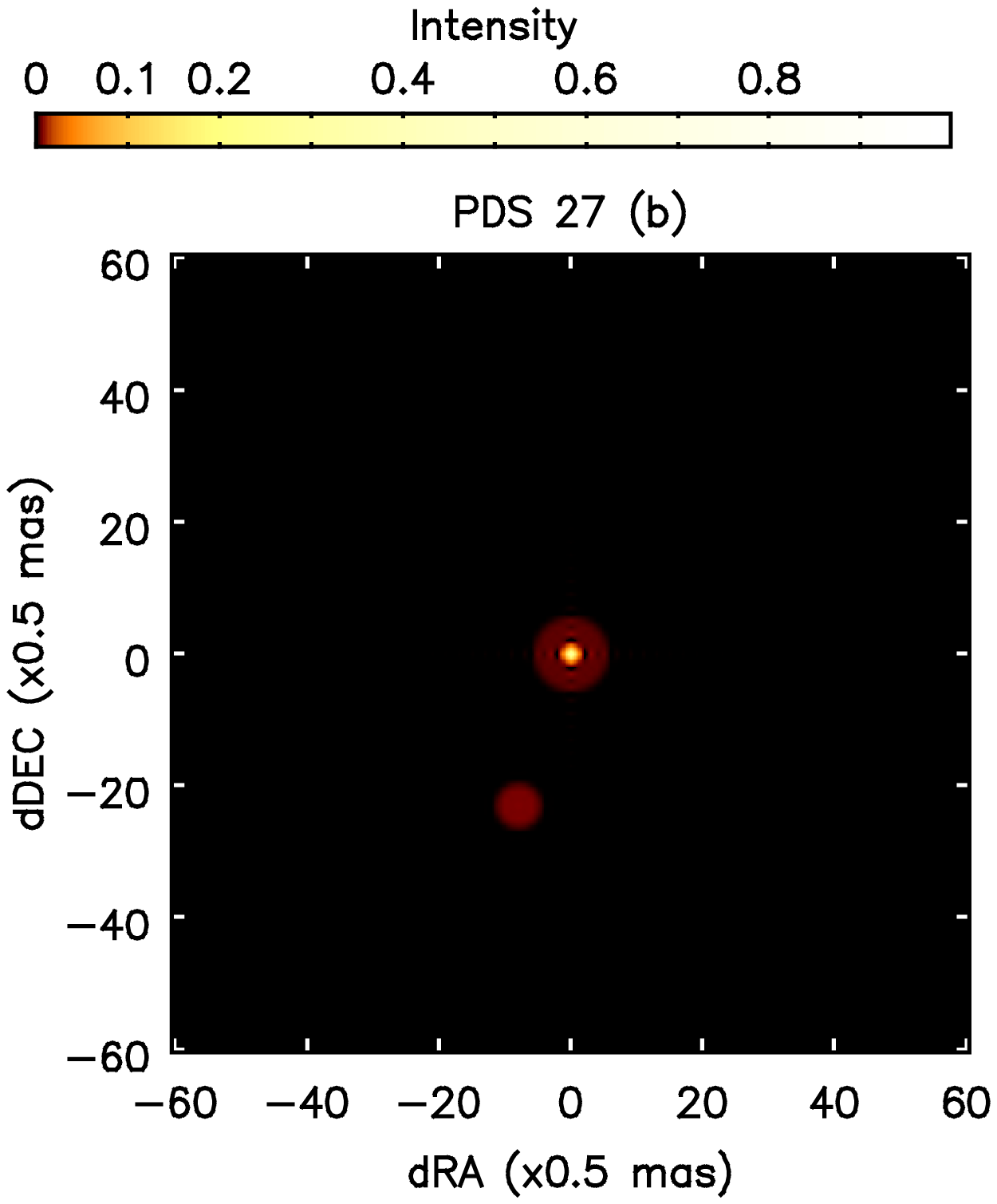} &
\includegraphics[scale=0.4,trim={0 0 0 0},clip=true]{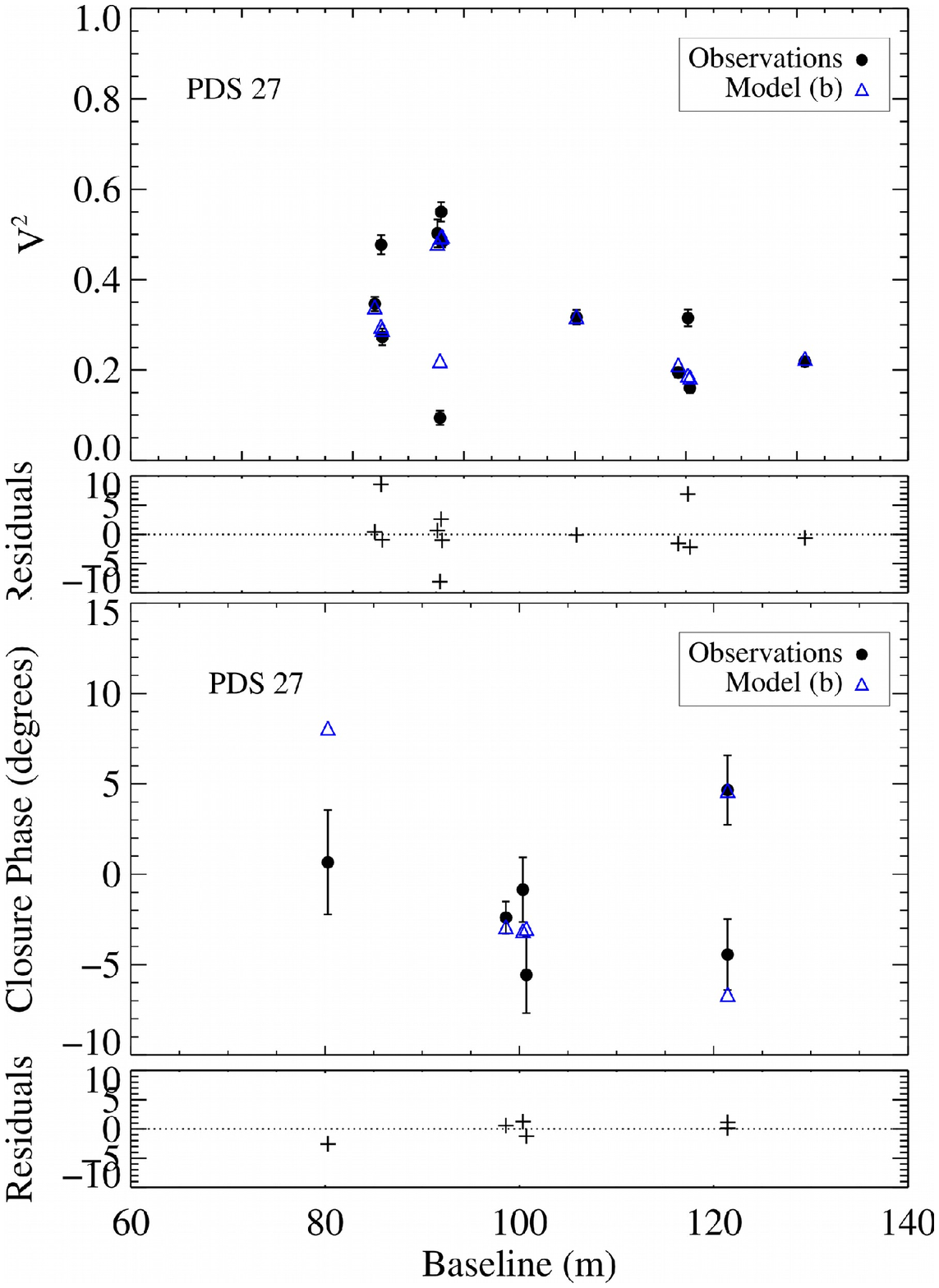} \\
\end{array}$
\end{center}
          \caption{Best-fit geometrical models (a) flattened ring (Red. ${\chi}^2$ $\sim$10) and b) binary (Red. ${\chi}^2$ $\sim$12)) are compared with the observed visibilities (V$^{2}$) and closure phases of PDS 27. The black points with vertical error bars are the observed data, and the blue triangles correspond to a representative best-fit model. The two models show a similar fit, and therefore we show model b as a representative fit.}
   \label{models_27}
\end{figure*}

\vskip 0.1in
{\bf{PDS 27.}} For this object, a similar procedure was followed as for PDS 37. In Fig.~\ref{models_27} we present two different models that provide a good fit for both visibilities and closure phases: a) a flattened ring surrounding a central object (${\chi}^2$ $\sim$ 10) and b) a binary with two resolved components (${\chi}^2$ $\sim$ 12). In the case of the flattened ring, the best-fit results in a flux ratio of 1 (ring over central object), an inner diameter of 2.6 mas, and a PA (minor axis) of 12.6$^{\circ}$. In the case of a binary with two resolved components, the best-fit results in a flux ratio of 0.2$\pm$0.2 (secondary/primary), a primary disk size of 5.4 mas, and a secondary disk size of 3 mas. The secondary companion is located at a separation of $\sim$ 12 mas (PA $\sim$ 200$^{\circ}$), which corresponds to 30 au at the distance of PDS 27. The resulting parameters of the two models with their errors are presented in Table~\ref{results2}. The uncertainties of flux ratios and exact positions of the secondary in our models are high. In particular, the position of the secondary shows several local minima that make the solution less robust. 

The interferometric results in combination with the radial velocity analysis in Sec.~\ref{spec} favor the scenario that PDS 27 is likely a binary system, with probably two resolved disks. The interferometric models alone cannot exclude the flattened-ring scenario (similar best fit ${\chi}^2$). For a better characterization of the system, more interferometric data are required.

\subsection{Radial velocity analysis}

\label{spec}

For PDS 27 we used the radial velocity measurements at four different epochs, and we proceed with the analysis of a single-lined spectroscopic binary. To do so, we used the rvfit code \citep{Iglesias2015}, assuming a fixed eccentricity of 0 and an inclination i=90$^{\circ}$ (edge-on). Since the actual inclination of the system is unknown, the radial velocities and the derived masses for i=90$^{\circ}$ provide a lower bound on the actual parameters of the system. In our approach the period and time of periastron passage of the system were set as free parameters at all times, while the systemic velocity and the amplitude of the radial velocity were set free only one at a time during fine-tuning so that the number of free parameters $\text{exceeds the}$ number of data points. During the fitting process we found a range of mass functions: 0.8~M$_\odot$ -- 1.7~M$_\odot$ corresponding to 6.7~M$_\odot$ -- 8.7~M$_\odot$ for the secondary object, periods of 5.5 -- 12 yr, and separations ($\alpha$=$\alpha_1$M$_{tot}$/M$_{2}$) of 9 -- 15 au. For these calculations we assumed that the observed velocity shifts correspond to the brightest, more massive object (M$_{1}$=12~M$_\odot$). The modeled separation is smaller by at least a factor of 2 than the separation we derived from the interferometry (30 au), but the two values are on the same order of magnitude, giving a good estimate of the actual separation of the binary. Based on the consistency of the independent datasets, we conclude that it is most likely that PDS 27 is a close MYSO binary. In the case of PDS 37, no spectroscopic velocity offset was observed in a period of two years, and therefore we did not perform a similar analysis.  

\section{Discussion}
\label{c3:sec:dis}

Overall, the interferometric results indicate a binary nature of the two objects. In particular, the observed oscillations in the visibility and the observed non-zero closure phase, especially toward PDS 37, point toward asymmetries and geometries that can best be explained when a binary source is assumed. The scenario of an asymmetric disk or ring in H band would need it to be extremely azimuthally modulated to reproduce the high-contrast visibilities we observe. No such required degree of modulation is observed in HAeBes with PIONIER \citep[see, e.g.,][]{Lazareff2017}. In addition, although our interferometric results point more toward the binary nature of PDS 37, previous spectroscopic and spectropolarimetric results \citep{Ilee2013,Ababakr2015} indicated a flattened active circumstellar disk seen close to edge-on. Our results on PDS 27 are supported by previous spectroscopic velocity measurements. It appears that the binary model provides the best fit, showing signatures of extended emission that need further investigation (see Appendix~\ref{smear}). Our data cannot constrain all the fitting parameters because there are degeneracies. We find two out of two objects \citep[100\%, in agreement with][]{Pomohaci2019} to be binaries with small separations (30~au for PDS 27; 42-54~au for PDS 37). 

Assuming that the extended H-band emission is due to hot dust, the angular sizes can be estimated. The PIONIER wavelength range (1.533\,$\mu$m - 1.772\,$\mu$m) is sensitive to thermal radiation of the dust in the temperature range 1500 to 2000 K. The angular distance of the dust from the star can then be estimated using d = $\frac{R\star}{2}\frac{T\star^{2}}{T_d^{2}}$, where $d$ is the distance of the dust from the star, $R\star$ and $T\star$ are the radius and temperature of the star, and $T_{d}$ is the dust temperature. The stellar radii of PDS 37 and PDS 27 are $\sim$13.7~${\rm R}_\odot$ and $\sim$11.7~${\rm R}_\odot$ , respectively \citep{Vioque2018}, while their temperature is $\sim$17500~K \citep{Ababakr2015}. For a dust temperature of 1500~K (graphite grains), an angular dust sublimation radius of $\sim$ 3.2 au (1.7 mas at 1.93 kpc) is estimated for PDS 37 and $\sim$ 3.8 au (1.5 mas at 2.55 kpc) for PDS 27. The maximum achieved angular resolution with PIONIER (1.3 mas) at the measured distance of our objects is $\sim$ 3.3 au and 2.5 au for PDS 27 and PDS 37, respectively, which would correspond to the very inner parts of the circumstellar disk or ring. In the case of PDS 27, the best-fit ring model resulted in an inner radius of 1.3$\pm$0.2 mas, which is in agreement with the dust sublimation radius of the object (1.5 mas). In the case of PDS 37, the best-fit ring model resulted in an inner radius of 3.8$\pm$0.2 mas, which is larger by a factor of 2 than the dust sublimation radius (1.7 mas). This difference could be explained with the presence of a particularly small dust-grain population, by inhomogeneities of the circumstellar environment, gaps or holes in the inner regions of the disks as part of evolution, or that the inferred temperature has been underestimated. We conclude that our ring models are consistent with the expected dust sublimation radius, and that the instrument resolution allows us to trace these inner regions, which are excavated of dust.

In the case of PDS 27, the secondary central object is absent and the flux of the secondary disk is only 22\% ($\pm$ 20\%) of the flux of the primary. Therefore, the secondary object is a deeply embedded lower-mass companion. This fact in combination with the small separation of $\sim$30$\pm$17~au and a predicted period of $\sim$ 10 years places PDS 27 among the very few observed candidates to directly test the theories of young binary formation through accretion disk fragmentation \citep{Meyer2018}. Our finding might support the theoretical predictions of the disk fragmentation channel as a possible mechanism for the formation of close MYSO binaries ($<$10 au), having a high-mass and an accreting low-mass component. Such systems are expected to be the progenitors of the short-period ($<$10 d) massive spectroscopic binaries. 

Future interferometric observations (e.g., VLTI) of PDS 27 and PDS 37 will provide a more complete uv-plane, which is necessary for the better characterization of the systems. The additional data will make it possible to constrain the circumstellar and circumbinary environment of the two components of the system and test the coplanarity of circumbinary disks around close binaries, evidence that might provide a further test for the disk fragmentation origin of such systems \citep{Duchene2015}.

\section{Summary and conclusions}

We reported the discovery of two massive (M$>$8M$_\odot$) YSO binaries at very close separations. Our main findings are listed below.

\begin{itemize}

\item The PIONIER data are consistent with both PDS 27 and PDS 37 being MYSO binaries. The binary nature of PDS 27 is also supported by spectroscopic observations. The observed velocity offsets in multiple epoch {\it V$_{\rm LSR}$} suggest a binary system with a period of $\sim$ 10 years.   

\item We find small component separations of 42-54 au towards PDS 37 and 30 au towards PDS 27. The closest separation of MYSO binaries that were spatially resolved so far is 30 au.

\end{itemize}

%
  
\begin{acknowledgements}

We would like to thank the anonymous referee for providing helpful comments and suggestions that improved the paper.
S.K. acknowledges support from an ERC Starting Grant (Grant Agreement No.\ 639889). A.C.G. has received funding from the European Research Council (ERC) under the European Union's Horizon 2020 research and innovation programme (grant agreement No.\ 743029).
      
\end{acknowledgements}

%
\bibliographystyle{aa.bst} 
\bibliography{references.bib} 

\begin{thebibliography}{49}
\expandafter\ifx\csname natexlab\endcsname\relax\def\natexlab#1{#1}\fi

\bibitem[{{Ababakr} {et~al.}(2015){Ababakr}, {Fairlamb}, {Oudmaijer}, \& {van
  den Ancker}}]{Ababakr2015}
{Ababakr}, K.~M., {Fairlamb}, J.~R., {Oudmaijer}, R.~D., \& {van den Ancker},
  M.~E. 2015, \mnras, 452, 2566

\bibitem[{{Baines} {et~al.}(2006){Baines}, {Oudmaijer}, {Porter}, \&
  {Pozzo}}]{Baines2006}
{Baines}, D., {Oudmaijer}, R.~D., {Porter}, J.~M., \& {Pozzo}, M. 2006, \mnras,
  367, 737

\bibitem[{{Bate} {et~al.}(2002){Bate}, {Bonnell}, \& {Bromm}}]{Bate2002}
{Bate}, M.~R., {Bonnell}, I.~A., \& {Bromm}, V. 2002, \mnras, 336, 705

\bibitem[{{Beltr{\'a}n} \& {de Wit}(2016)}]{Beltran2016}
{Beltr{\'a}n}, M.~T. \& {de Wit}, W.~J. 2016, \aapr, 24, 6

\bibitem[{{Beuther} {et~al.}(2017){Beuther}, {Linz}, {Henning}, {Feng}, \&
  {Teague}}]{Beuther2017}
{Beuther}, H., {Linz}, H., {Henning}, T., {Feng}, S., \& {Teague}, R. 2017,
  \aap, 605, A61

\bibitem[{{Bonneau} {et~al.}(2011){Bonneau}, {Delfosse}, {Mourard}, {Lafrasse},
  {Mella}, {Cetre}, {Clausse}, \& {Zins}}]{Bonneau2011}
{Bonneau}, D., {Delfosse}, X., {Mourard}, D., {et~al.} 2011, \aap, 535, A53

\bibitem[{{Bonnell} \& {Bate}(2005)}]{Bonnell2005}
{Bonnell}, I.~A. \& {Bate}, M.~R. 2005, \mnras, 362, 915

\bibitem[{{Chini} {et~al.}(2012){Chini}, {Hoffmeister}, {Nasseri}, {Stahl}, \&
  {Zinnecker}}]{Chini2012}
{Chini}, R., {Hoffmeister}, V.~H., {Nasseri}, A., {Stahl}, O., \& {Zinnecker},
  H. 2012, \mnras, 424, 1925

\bibitem[{{Davis} {et~al.}(2000){Davis}, {Tango}, \& {Booth}}]{Davis2000}
{Davis}, J., {Tango}, W.~J., \& {Booth}, A.~J. 2000, \mnras, 318, 387

\bibitem[{{Duch{\^e}ne}(2015)}]{Duchene2015}
{Duch{\^e}ne}, G. 2015, \apss, 355, 291

\bibitem[{{Hartmann} {et~al.}(2016){Hartmann}, {Herczeg}, \&
  {Calvet}}]{Hartmann2016}
{Hartmann}, L., {Herczeg}, G., \& {Calvet}, N. 2016, \araa, 54, 135

\bibitem[{{Iglesias-Marzoa} {et~al.}(2015){Iglesias-Marzoa},
  {L{\'o}pez-Morales}, \& {Jes{\'u}s Ar{\'e}valo Morales}}]{Iglesias2015}
{Iglesias-Marzoa}, R., {L{\'o}pez-Morales}, M., \& {Jes{\'u}s Ar{\'e}valo
  Morales}, M. 2015, \pasp, 127, 567

\bibitem[{{Ilee} {et~al.}(2016){Ilee}, {Cyganowski}, {Nazari}, {Hunter},
  {Brogan}, {Forgan}, \& {Zhang}}]{Ilee2016}
{Ilee}, J.~D., {Cyganowski}, C.~J., {Nazari}, P., {et~al.} 2016, \mnras, 462,
  4386

\bibitem[{{Ilee} {et~al.}(2018){Ilee}, {Oudmaijer}, {Wheelwright}, \&
  {Pomohaci}}]{Ilee2018}
{Ilee}, J.~D., {Oudmaijer}, R.~D., {Wheelwright}, H.~E., \& {Pomohaci}, R.
  2018, \mnras, 477, 3360

\bibitem[{{Ilee} {et~al.}(2013){Ilee}, {Wheelwright}, {Oudmaijer}, {de Wit},
  {Maud}, {Hoare}, {Lumsden}, {Moore}, {Urquhart}, \& {Mottram}}]{Ilee2013}
{Ilee}, J.~D., {Wheelwright}, H.~E., {Oudmaijer}, R.~D., {et~al.} 2013, \mnras,
  429, 2960

\bibitem[{{Jijina} \& {Adams}(1996)}]{Jijina1996}
{Jijina}, J. \& {Adams}, F.~C. 1996, \apj, 462, 874

\bibitem[{{Johnston} {et~al.}(2015){Johnston}, {Robitaille}, {Beuther}, {Linz},
  {Boley}, {Kuiper}, {Keto}, {Hoare}, \& {van Boekel}}]{Johnston2015}
{Johnston}, K.~G., {Robitaille}, T.~P., {Beuther}, H., {et~al.} 2015, \apjl,
  813, L19

\bibitem[{{Karl} {et~al.}(2018){Karl}, {Pfuhl}, {Eisenhauer}, {Genzel},
  {Grellmann}, {Habibi}, {Abuter}, {Accardo}, {Amorim}, {Anugu}, {{\'A}vila},
  {Benisty}, {Berger}, {Bland}, {Bonnet}, {Bourget}, {Brandner}, {Brast},
  {Buron}, {Garatti}, {Chapron}, {Cl{\'e}net}, {Collin}, {Coud{\'e} du
  Foresto}, {de Wit}, {de Zeeuw}, {Deen}, {Delplancke-Str{\"o}bele}, {Dembet},
  {Derie}, {Dexter}, {Duvert}, {Ebert}, {Eckart}, {Esselborn}, {F{\'e}dou},
  {Finger}, {Garcia}, {Garcia Dabo}, {Garcia Lopez}, {Gao}, {Gandron},
  {Gillessen}, {Gont{\'e}}, {Gordo}, {Gr{\"o}zinger}, {Guajardo}, {Guieu},
  {Haguenauer}, {Hans}, {Haubois}, {Haug}, {Hau{\ss}mann}, {Henning},
  {Hippler}, {Horrobin}, {Huber}, {Hubert}, {Hubin}, {Hummel}, {Jakob},
  {Jochum}, {Jocou}, {Kaufer}, {Kellner}, {Kandrew}, {Kern}, {Kervella},
  {Kiekebusch}, {Klein}, {K{\"o}hler}, {Kolb}, {Kulas}, {Lacour},
  {Lapeyr{\`e}re}, {Lazareff}, {Le Bouquin}, {L{\'e}na}, {Lenzen},
  {L{\'e}v{\^e}que}, {Lin}, {Lippa}, {Magnard}, {Mehrgan}, {M{\'e}rand},
  {Moulin}, {M{\"u}ller}, {M{\"u}ller}, {Neumann}, {Oberti}, {Ott}, {Pallanca},
  {Panduro}, {Pasquini}, {Paumard}, {Percheron}, {Perraut}, {Perrin},
  {Pfl{\"u}ger}, {Phan Duc}, {Plewa}, {Popovic}, {Rabien}, {Ram{\'{\i}}rez},
  {Ramos}, {Rau}, {Riquelme}, {Rodr{\'{\i}}guez-Coira}, {Rohloff}, {Rosales},
  {Rousset}, {Sanchez-Bermudez}, {Scheithauer}, {Sch{\"o}ller}, {Schuhler},
  {Spyromilio}, {Straub}, {Straubmeier}, {Sturm}, {Suarez}, {Tristram},
  {Ventura}, {Vincent}, {Waisberg}, {Wank}, {Widmann}, {Wieprecht}, {Wiest},
  {Wiezorrek}, {Wittkowski}, {Woillez}, {Wolff}, {Yazici}, {Ziegler}, \&
  {Zins}}]{Karl2018}
{Karl}, M., {Pfuhl}, O., {Eisenhauer}, F., {et~al.} 2018, ArXiv e-prints
  [\eprint[arXiv]{1809.10376}]

\bibitem[{{Klassen} {et~al.}(2016){Klassen}, {Pudritz}, {Kuiper}, {Peters}, \&
  {Banerjee}}]{Klassen2016}
{Klassen}, M., {Pudritz}, R.~E., {Kuiper}, R., {Peters}, T., \& {Banerjee}, R.
  2016, \apj, 823, 28

\bibitem[{{Kraus} {et~al.}(2006){Kraus}, {Balega}, {Elitzur}, {Hofmann},
  {Preibisch}, {Rosen}, {Schertl}, {Weigelt}, \& {Young}}]{Kraus2006}
{Kraus}, S., {Balega}, Y., {Elitzur}, M., {et~al.} 2006, \aap, 455, 521

\bibitem[{{Kraus} {et~al.}(2012){Kraus}, {Calvet}, {Hartmann}, {Hofmann},
  {Kreplin}, {Monnier}, \& {Weigelt}}]{Kraus2012}
{Kraus}, S., {Calvet}, N., {Hartmann}, L., {et~al.} 2012, \apj, 752, 11

\bibitem[{{Kraus} {et~al.}(2010){Kraus}, {Hofmann}, {Menten}, {Schertl},
  {Weigelt}, {Wyrowski}, {Meilland}, {Perraut}, {Petrov}, {Robbe-Dubois},
  {Schilke}, \& {Testi}}]{Kraus2010}
{Kraus}, S., {Hofmann}, K.-H., {Menten}, K.~M., {et~al.} 2010, \nat, 466, 339

\bibitem[{{Kraus} {et~al.}(2017){Kraus}, {Kluska}, {Kreplin}, {Bate},
  {Harries}, {Hofmann}, {Hone}, {Monnier}, {Weigelt}, {Anugu}, {de Wit}, \&
  {Wittkowski}}]{Kraus2017}
{Kraus}, S., {Kluska}, J., {Kreplin}, A., {et~al.} 2017, \apjl, 835, L5

\bibitem[{{Kraus} {et~al.}(2005){Kraus}, {Schloerb}, {Traub}, {Carleton},
  {Lacasse}, {Pearlman}, {Monnier}, {Millan-Gabet}, {Berger}, {Haguenauer},
  {Perraut}, {Kern}, {Malbet}, \& {Labeye}}]{Kraus2005}
{Kraus}, S., {Schloerb}, F.~P., {Traub}, W.~A., {et~al.} 2005, \aj, 130, 246

\bibitem[{{Krumholz} {et~al.}(2012){Krumholz}, {Klein}, \&
  {McKee}}]{Krumholz2012}
{Krumholz}, M.~R., {Klein}, R.~I., \& {McKee}, C.~F. 2012, \apj, 754, 71

\bibitem[{{Krumholz} {et~al.}(2009){Krumholz}, {Klein}, {McKee}, {Offner}, \&
  {Cunningham}}]{Krumholz2009}
{Krumholz}, M.~R., {Klein}, R.~I., {McKee}, C.~F., {Offner}, S.~S.~R., \&
  {Cunningham}, A.~J. 2009, Science, 323, 754

\bibitem[{{Kuiper} \& {Hosokawa}(2018)}]{Kuiper2018}
{Kuiper}, R. \& {Hosokawa}, T. 2018, ArXiv e-prints
  [\eprint[arXiv]{1804.10211}]

\bibitem[{{Kuiper} {et~al.}(2010){Kuiper}, {Klahr}, {Beuther}, \&
  {Henning}}]{kuiper10}
{Kuiper}, R., {Klahr}, H., {Beuther}, H., \& {Henning}, T. 2010, \apj, 722,
  1556

\bibitem[{{Lachaume} \& {Berger}(2013)}]{Lachaume2013}
{Lachaume}, R. \& {Berger}, J.-P. 2013, \mnras, 435, 2501

\bibitem[{{Lazareff} {et~al.}(2017){Lazareff}, {Berger}, {Kluska}, {Le
  Bouquin}, {Benisty}, {Malbet}, {Koen}, {Pinte}, {Thi}, {Absil}, {Baron},
  {Delboulb{\'e}}, {Duvert}, {Isella}, {Jocou}, {Juhasz}, {Kraus}, {Lachaume},
  {M{\'e}nard}, {Millan-Gabet}, {Monnier}, {Moulin}, {Perraut}, {Rochat},
  {Soulez}, {Tallon}, {Thi{\'e}baut}, {Traub}, \& {Zins}}]{Lazareff2017}
{Lazareff}, B., {Berger}, J.-P., {Kluska}, J., {et~al.} 2017, \aap, 599, A85

\bibitem[{{Le Bouquin} {et~al.}(2011){Le Bouquin}, {Berger}, {Lazareff},
  {Zins}, {Haguenauer}, {Jocou}, {Kern}, {Millan-Gabet}, {Traub}, {Absil},
  {Augereau}, {Benisty}, {Blind}, {Bonfils}, {Bourget}, {Delboulbe},
  {Feautrier}, {Germain}, {Gitton}, {Gillier}, {Kiekebusch}, {Kluska},
  {Knudstrup}, {Labeye}, {Lizon}, {Monin}, {Magnard}, {Malbet}, {Maurel},
  {M{\'e}nard}, {Micallef}, {Michaud}, {Montagnier}, {Morel}, {Moulin},
  {Perraut}, {Popovic}, {Rabou}, {Rochat}, {Rojas}, {Roussel}, {Roux},
  {Stadler}, {Stefl}, {Tatulli}, \& {Ventura}}]{LeBouquin2011}
{Le Bouquin}, J.-B., {Berger}, J.-P., {Lazareff}, B., {et~al.} 2011, \aap, 535,
  A67

\bibitem[{{Lumsden} {et~al.}(2013){Lumsden}, {Hoare}, {Urquhart}, {Oudmaijer},
  {Davies}, {Mottram}, {Cooper}, \& {Moore}}]{Lumsden2013}
{Lumsden}, S.~L., {Hoare}, M.~G., {Urquhart}, J.~S., {et~al.} 2013, \apjs, 208,
  11

\bibitem[{{Lund} \& {Bonnell}(2018)}]{Lund2018}
{Lund}, K. \& {Bonnell}, I.~A. 2018, \mnras [\eprint[arXiv]{1806.07394}]

\bibitem[{{Meyer} {et~al.}(2018){Meyer}, {Kuiper}, {Kley}, {Johnston}, \&
  {Vorobyov}}]{Meyer2018}
{Meyer}, D.~M.-A., {Kuiper}, R., {Kley}, W., {Johnston}, K.~G., \& {Vorobyov},
  E. 2018, \mnras, 473, 3615

\bibitem[{{Moe} \& {Di Stefano}(2017)}]{Moe2017}
{Moe}, M. \& {Di Stefano}, R. 2017, \apjs, 230, 15

\bibitem[{{Monnier}(2003)}]{Monnier2003}
{Monnier}, J.~D. 2003, in EAS Publications Series, Vol.~6, EAS Publications
  Series, ed. G.~{Perrin} \& F.~{Malbet}, 213

\bibitem[{{Myers} {et~al.}(2013){Myers}, {McKee}, {Cunningham}, {Klein}, \&
  {Krumholz}}]{Myers2013}
{Myers}, A.~T., {McKee}, C.~F., {Cunningham}, A.~J., {Klein}, R.~I., \&
  {Krumholz}, M.~R. 2013, \apj, 766, 97

\bibitem[{{Pomohaci} {et~al.}(2019){Pomohaci}, {Oudmaijer}, \&
  {Goodwin}}]{Pomohaci2019}
{Pomohaci}, R., {Oudmaijer}, R.~D., \& {Goodwin}, S.~P. 2019, \mnras, 484, 226

\bibitem[{{Rosen} {et~al.}(2016){Rosen}, {Krumholz}, {McKee}, \&
  {Klein}}]{Rosen2016}
{Rosen}, A.~L., {Krumholz}, M.~R., {McKee}, C.~F., \& {Klein}, R.~I. 2016,
  \mnras, 463, 2553

\bibitem[{{Sana} {et~al.}(2013){Sana}, {de Koter}, {de Mink}, {Dunstall},
  {Evans}, {H{\'e}nault-Brunet}, {Ma{\'{\i}}z Apell{\'a}niz},
  {Ram{\'{\i}}rez-Agudelo}, {Taylor}, {Walborn}, {Clark}, {Crowther},
  {Herrero}, {Gieles}, {Langer}, {Lennon}, \& {Vink}}]{Sana2013}
{Sana}, H., {de Koter}, A., {de Mink}, S.~E., {et~al.} 2013, \aap, 550, A107

\bibitem[{{Sana} {et~al.}(2012){Sana}, {de Mink}, {de Koter}, {Langer},
  {Evans}, {Gieles}, {Gosset}, {Izzard}, {Le Bouquin}, \&
  {Schneider}}]{Sana2012}
{Sana}, H., {de Mink}, S.~E., {de Koter}, A., {et~al.} 2012, Science, 337, 444

\bibitem[{{Sana} {et~al.}(2014){Sana}, {Le Bouquin}, {Lacour}, {Berger},
  {Duvert}, {Gauchet}, {Norris}, {Olofsson}, {Pickel}, {Zins}, {Absil}, {de
  Koter}, {Kratter}, {Schnurr}, \& {Zinnecker}}]{Sana2014}
{Sana}, H., {Le Bouquin}, J.-B., {Lacour}, S., {et~al.} 2014, \apjs, 215, 15

\bibitem[{{Szczerba} {et~al.}(2007){Szczerba}, {Si{\'o}dmiak}, {Stasi{\'n}ska},
  \& {Borkowski}}]{Szczerba2007}
{Szczerba}, R., {Si{\'o}dmiak}, N., {Stasi{\'n}ska}, G., \& {Borkowski}, J.
  2007, \aap, 469, 799

\bibitem[{{Tallon-Bosc} {et~al.}(2008){Tallon-Bosc}, {Tallon}, {Thi{\'e}baut},
  {B{\'e}chet}, {Mella}, {Lafrasse}, {Chesneau}, {Domiciano de Souza},
  {Duvert}, {Mourard}, {Petrov}, \& {Vannier}}]{Tallon2008}
{Tallon-Bosc}, I., {Tallon}, M., {Thi{\'e}baut}, E., {et~al.} 2008, in
  \procspie, Vol. 7013, Optical and Infrared Interferometry, 70131J

\bibitem[{{Vieira} {et~al.}(2011){Vieira}, {Gregorio-Hetem}, {Hetem},
  {Stasi{\'n}ska}, \& {Szczerba}}]{Vieira2011}
{Vieira}, R.~G., {Gregorio-Hetem}, J., {Hetem}, A., {Stasi{\'n}ska}, G., \&
  {Szczerba}, R. 2011, \aap, 526, A24

\bibitem[{{Vioque} {et~al.}(2018){Vioque}, {Oudmaijer}, {Baines},
  {Mendigut{\'{\i}}a}, \& {P{\'e}rez-Mart{\'{\i}}nez}}]{Vioque2018}
{Vioque}, M., {Oudmaijer}, R.~D., {Baines}, D., {Mendigut{\'{\i}}a}, I., \&
  {P{\'e}rez-Mart{\'{\i}}nez}, R. 2018, ArXiv e-prints
  [\eprint[arXiv]{1808.00476}]

\bibitem[{{Wheelwright} {et~al.}(2010){Wheelwright}, {Oudmaijer}, {de Wit},
  {Hoare}, {Lumsden}, \& {Urquhart}}]{Wheelwright2010}
{Wheelwright}, H.~E., {Oudmaijer}, R.~D., {de Wit}, W.~J., {et~al.} 2010,
  \mnras, 408, 1840

\bibitem[{{Wheelwright} {et~al.}(2011){Wheelwright}, {Vink}, {Oudmaijer}, \&
  {Drew}}]{Wheelwright2011}
{Wheelwright}, H.~E., {Vink}, J.~S., {Oudmaijer}, R.~D., \& {Drew}, J.~E. 2011,
  \aap, 532, A28

\bibitem[{{Yorke} \& {Sonnhalter}(2002)}]{Yorke2002}
{Yorke}, H.~W. \& {Sonnhalter}, C. 2002, \apj, 569, 846

\end{thebibliography}

\begin{appendix}

\section{Technical overview}

\begin{table}[ht]
\caption{Technical overview of the PIONIER observations of PDS 27 and PDS 37 over two nights in March 2015. The closure phases are the product of three baselines, and we report them at the highest spatial frequency (longest baseline of the triplet).}
\small
\centering
\setlength\tabcolsep{2pt}
\begin{tabular}{c c c c c c c}
\hline\hline
Source & MJD & Baseline & $\tau_{coh}$ & Seeing & V$^{2}$ & Closure Phase  \\ & & (m) & (ms) & (arcsec) & & ($^\circ$) \\
\hline\hline
PDS 27 & 57083.13889 & 44.0 & 6.7 & 0.83 & 0.35$\pm$0.02 & \\
&   & 45.2 &  &    & 0.48$\pm$0.02 &  \\
&   & 55.3  &  &    & 0.50$\pm$0.03 &  \\
&   & 56.1 &  &  & 0.55$\pm$0.02 &  \\
&   & 98.7  &  &    & 0.19$\pm$0.01 & -2.4$\pm$0.9 \\
&  & 100.4 & &   & 0.32$\pm$0.02 & -0.9$\pm$1.8 \\
& 57084.13284  & 45.3  & 5 & 0.93 & 0.27$\pm$0.01 & \\
&   & 55.7  &  &    & 0.09$\pm$0.01 &  \\
&   & 56.1  &  &    & 0.48$\pm$0.01 &  \\
&   & 80.3  &  &    & 0.32$\pm$0.02 & 0.7$\pm$2.9 \\
&   & 100.7  &  &    & 0.16$\pm$0.01 & -5.6$\pm$2.1 \\
&  & 121.4  & &  & 0.22$\pm$0.02 & -4.4$\pm$2.0/\\
&  &  &  &  & & 4.7$\pm$1.9\\
\hline
PDS 37 & 57084.18331 & 42.0 & 4.7 & 0.86 & 0.24$\pm$0.01 & \\
&   &  43.3 &  &  & 0.58$\pm$0.01 & \\
&   &  49.1 &  &  & 0.54$\pm$0.01 & \\
&   & 50.4 &  &  & 0.70$\pm$0.02 & \\
&   & 60.5 &  &  & 0.3$\pm$0.01 &  \\
&   &  61.8 &  &  & 0.09$\pm$0.01 & \\
&   &  88.9 &  &  & 0.38$\pm$0.02 & 0.2$\pm$2.0 \\
&   &  89.5 &  &  & 0.53$\pm$0.01 & 0.4$\pm$0.6 \\
&   &  90.3 &  &  & 0.14$\pm$0.01 & -1.0$\pm$2.5\\
&   &  93.0 &  &  & 0.56$\pm$0.03 & -1.0$\pm$3 \\
&   &  124.4 &  &  & 0.32$\pm$0.01 & 4.0$\pm$3.1/\\
&   &   &  &  &  & -0.7$\pm$2.8\\
&   &  127.4 &  &  & 0.45$\pm$0.01 & -0.3$\pm$2.4/ \\
&   &   &  &  &  & 1.5$\pm$1.8 \\
\hline\hline
\end{tabular}
\label{observables}
\end{table}

\begin{figure}[h]
\begin{center}
                \includegraphics[width=5cm,angle=90]{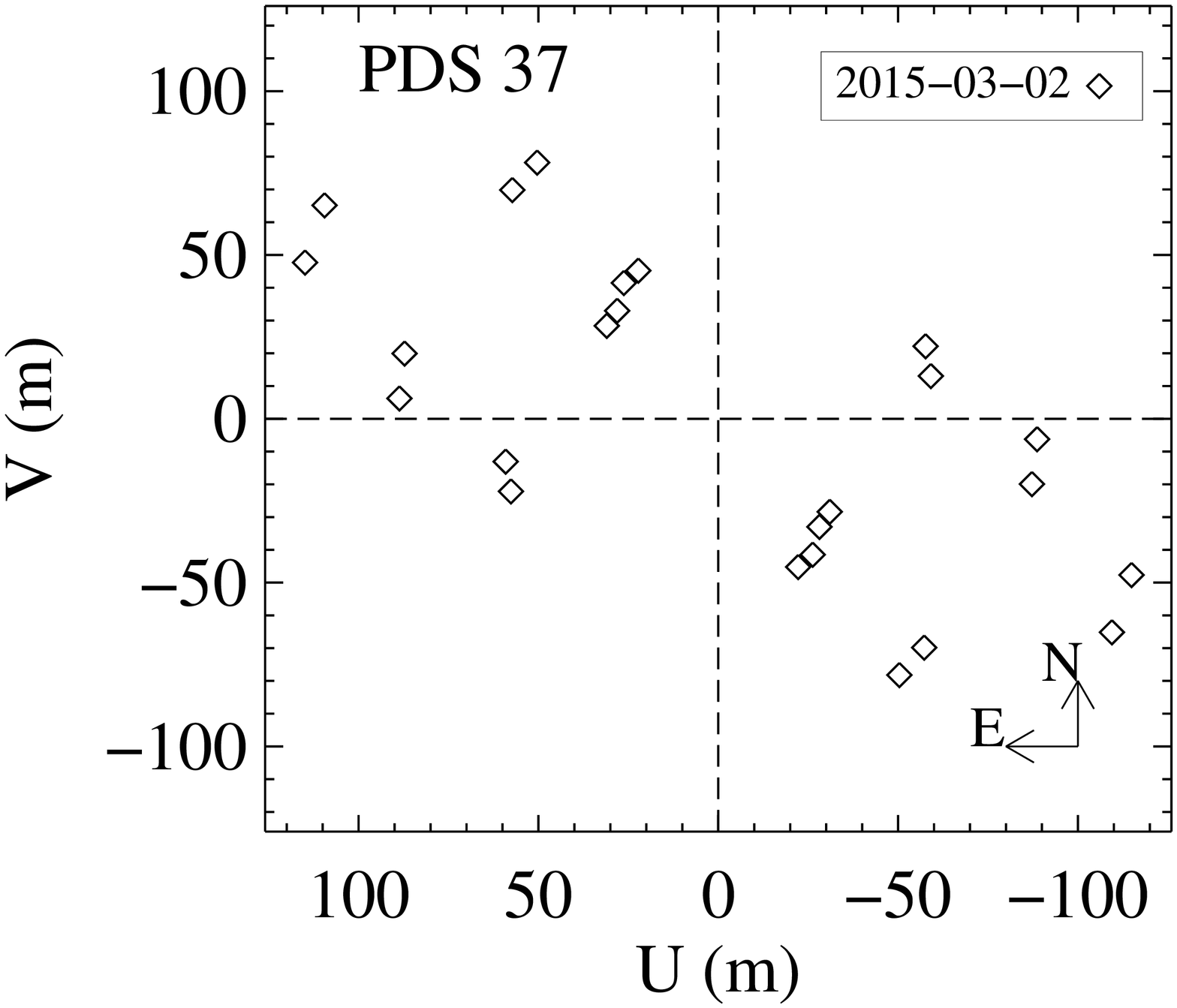} \\
                \includegraphics[width=5cm,angle=90]{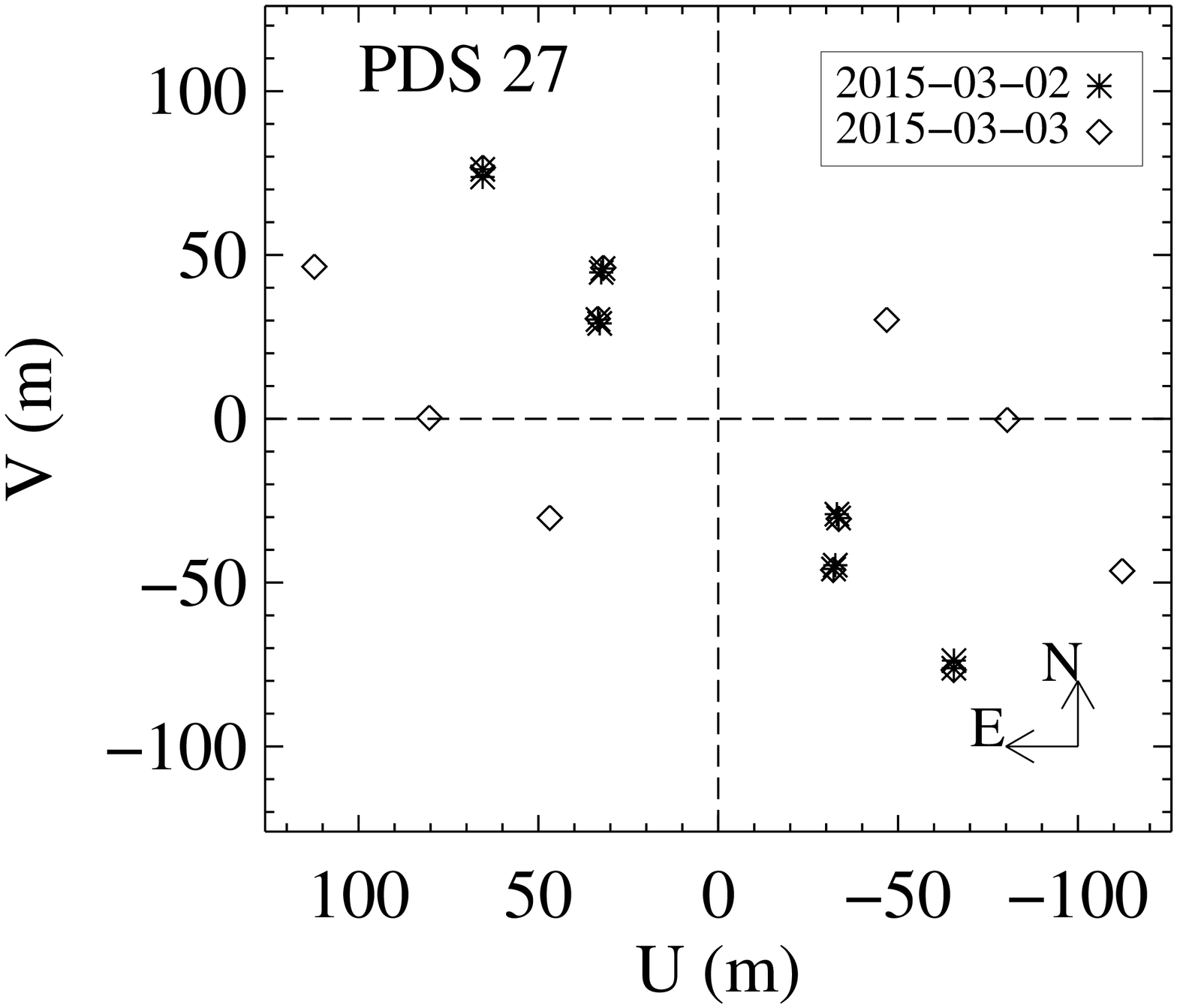} \\     
\end{center}
               \caption{PIONIER uv-coverage for PDS
                 37 (left) and PDS 27 (right).}
          \label{obs_uv}
          
\end{figure}  

\section{Fitting simple geometries}

\label{exploring}

The code LITpro computes modeled interferometric data for a given set of parameters and allows fitting the observed parameters by iteratively minimizing the residuals (difference between simulated and observed data). Because the interferometric models are not linear functions of the parameters, the parameter space is generally characterized by several local minima of $\chi^{2}$ for each of the parameters, and therefore finding a global solution is the main difficulty of the minimization process. Setting initial values of the fitting parameters and trying out different initial sets are both important parts of the fitting process. Following the optimization of a non-convex inverse problem, LITpro provides final chi-squares and error estimates on the estimated parameters that can be used to assess the quality of the fit. To minimize the risk of becoming trapped in a local minimum, we applied different initial ``seeds'' in the fitting engine of LITpro.    

The code LITpro offers a set of elementary geometric models (e.g., point source, Gaussian, disk, and ring), and the user has the flexibility to develop a more complex model by combining a subset of these. For instance, to model a binary of two unresolved stars, two point sources (Dirac functions) are used.   

\subsection{PDS 37}

\label{flux_dist_bin_1}

\begin{figure*}[h]
\begin{center}$ 
\begin{array}{ccc}
\includegraphics[width=4.4cm,angle=90]{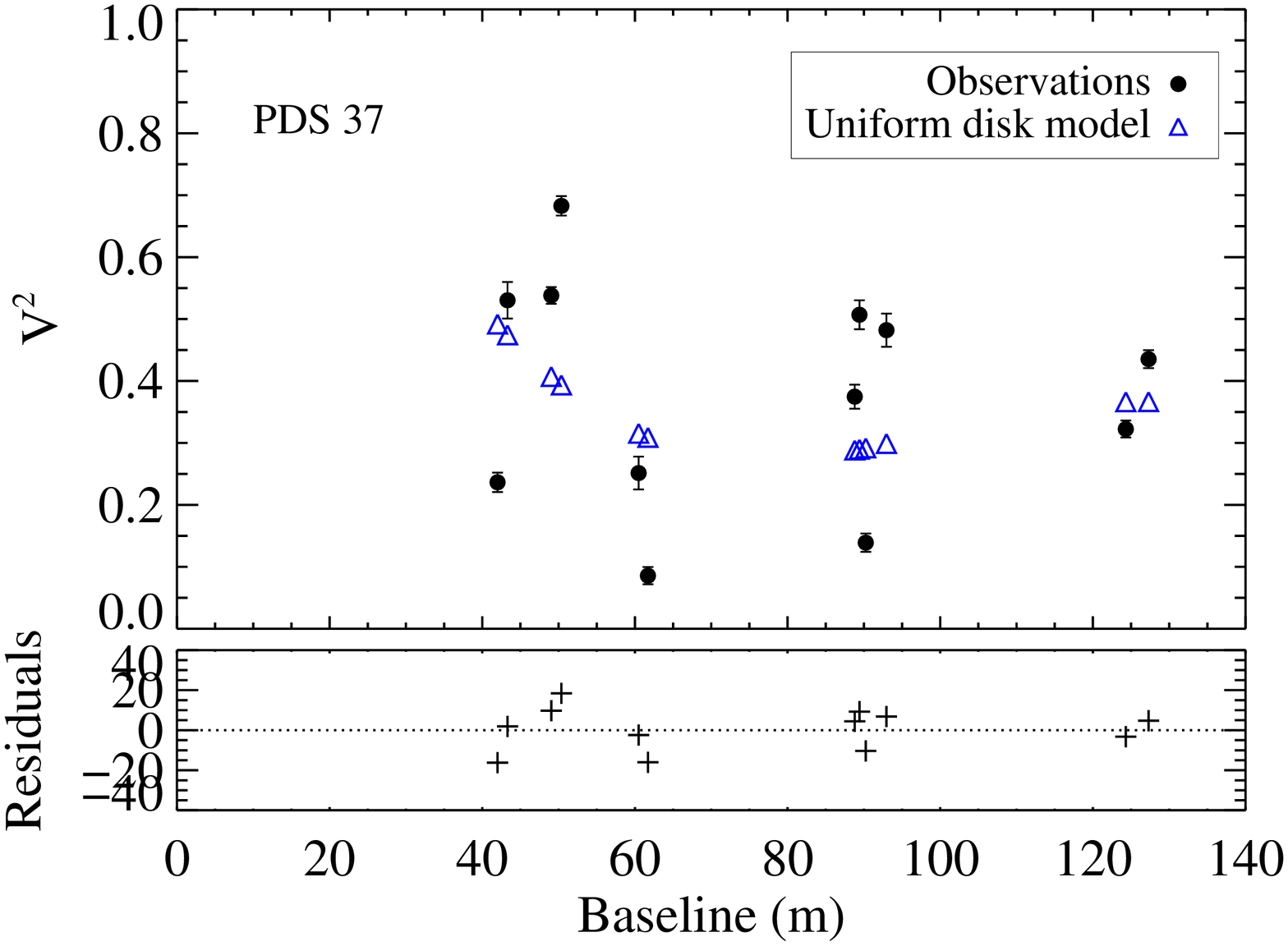} &
\includegraphics[width=4.4cm,angle=90]{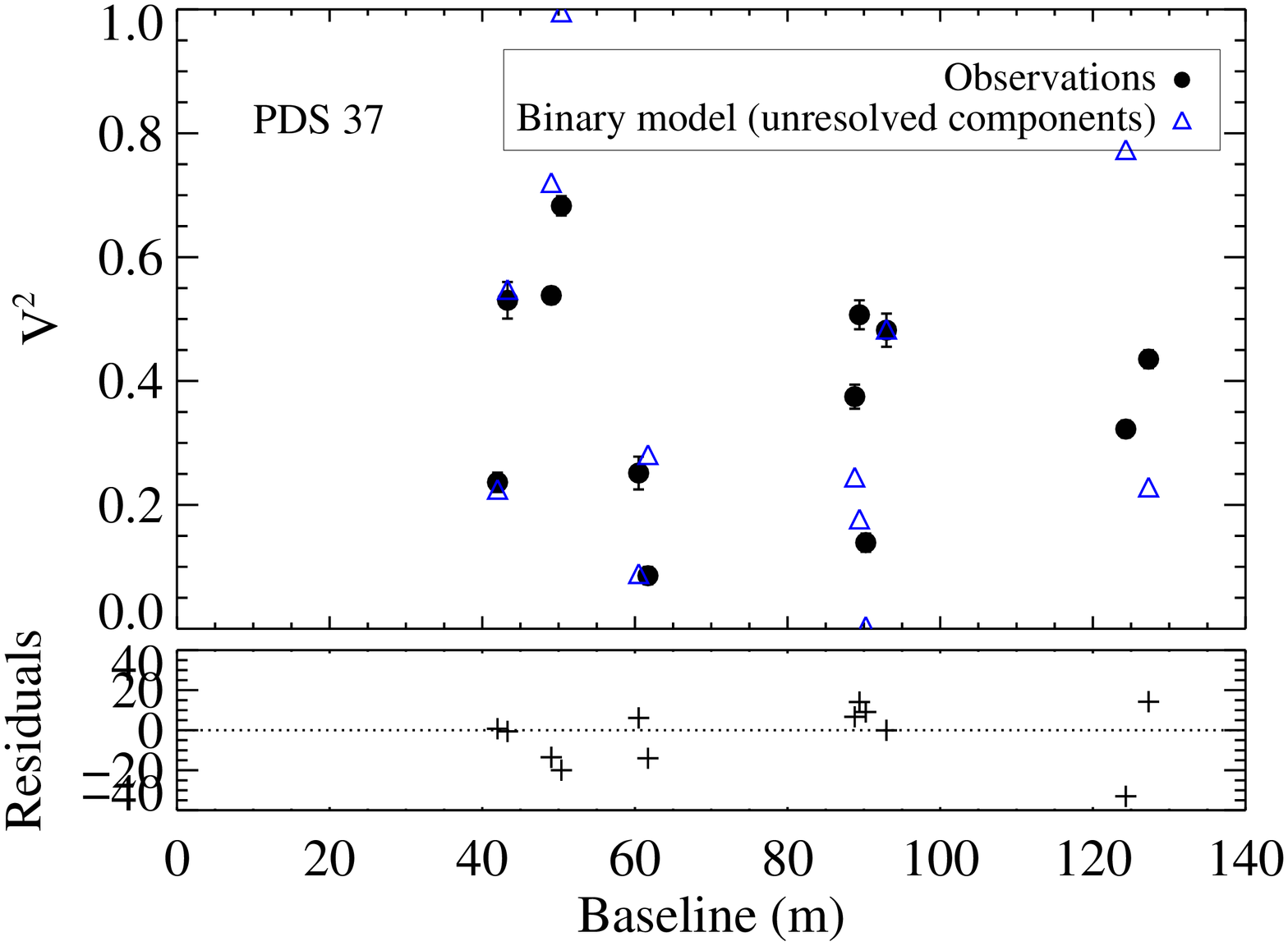} &
\includegraphics[width=4.4cm,angle=90]{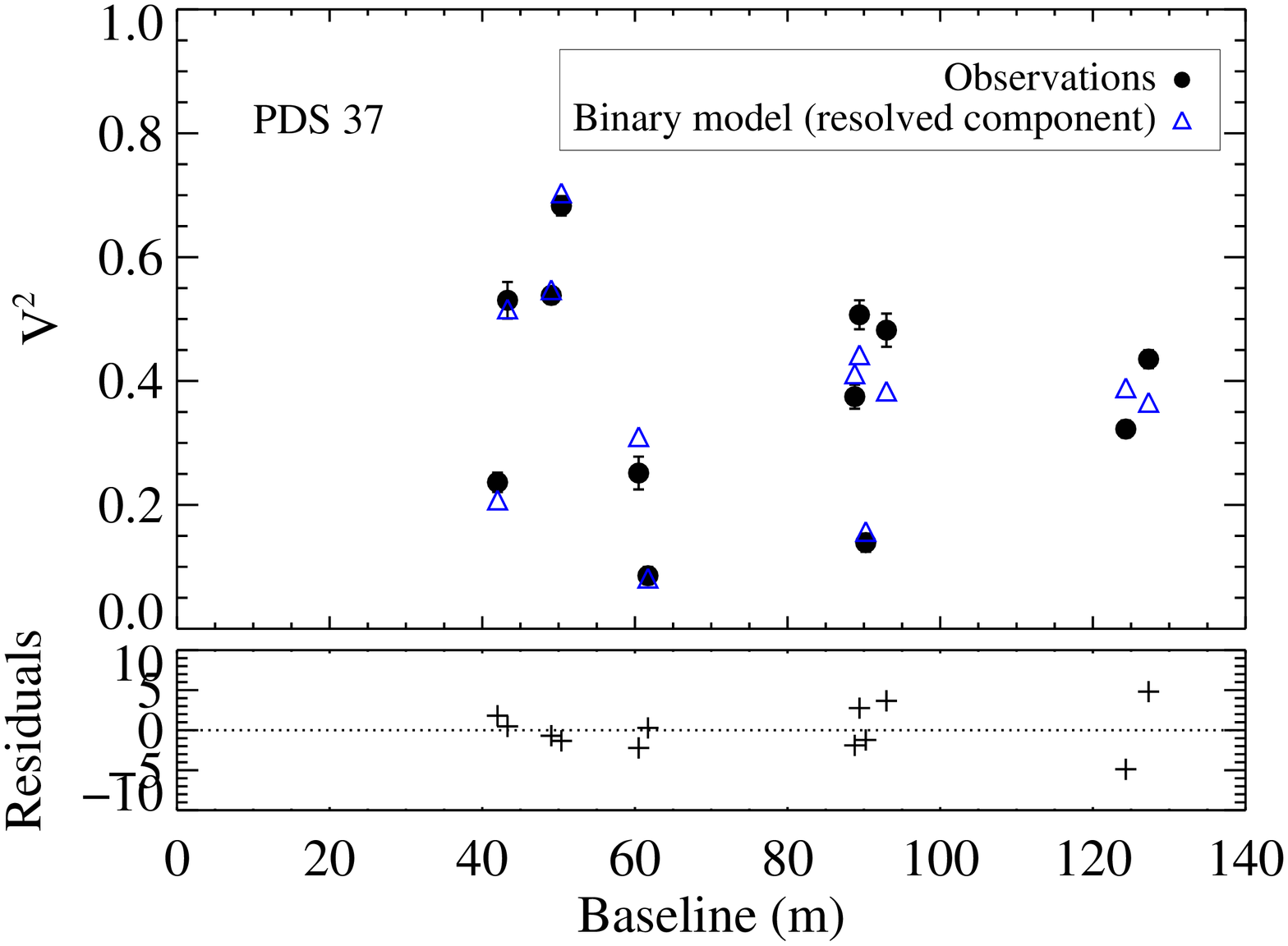} \\
\end{array}$
\end{center}
\centering
\includegraphics[width=4.4cm,angle=90]{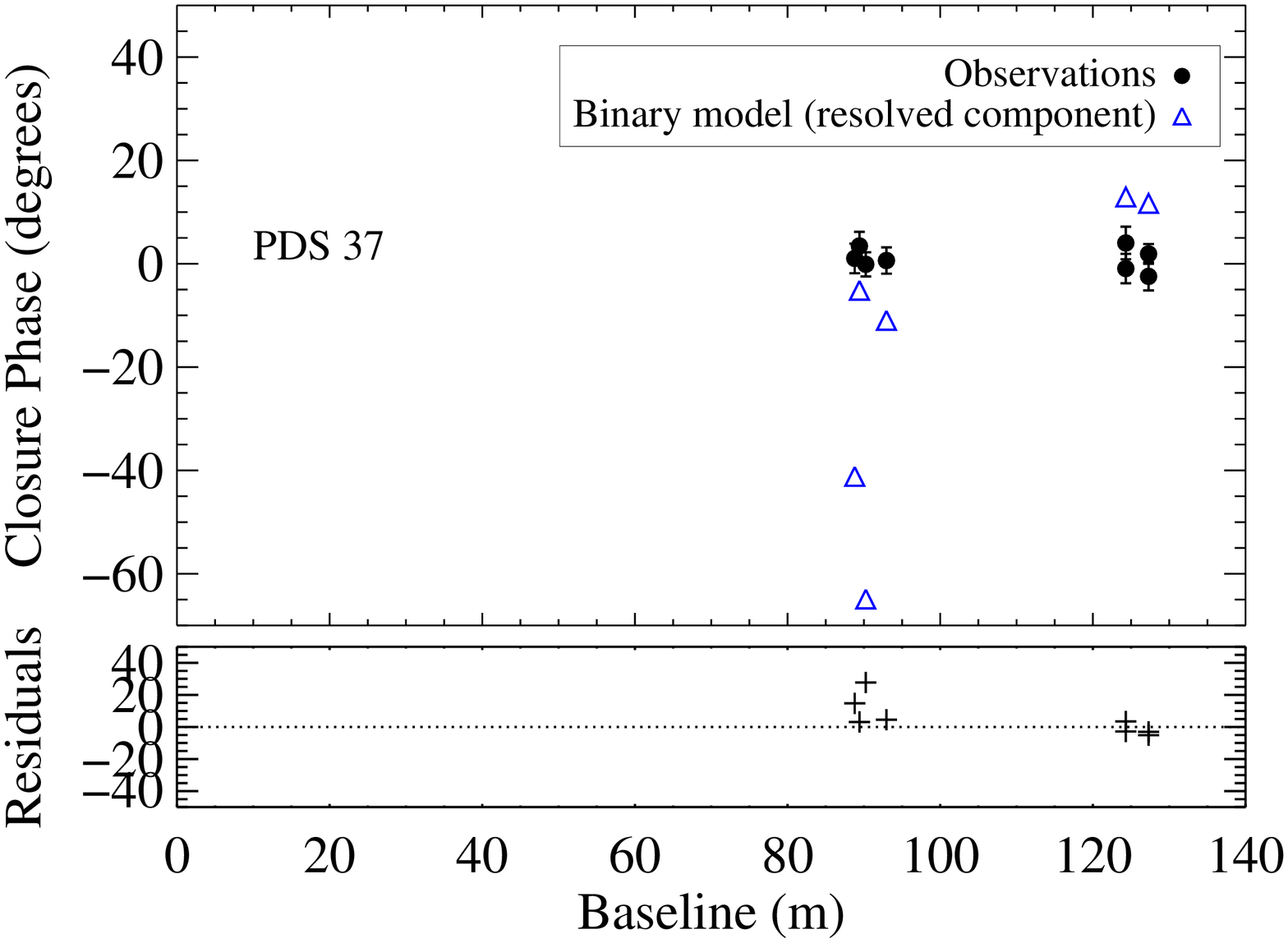}
\begin{center}$
\begin{array}{ccc}
                \includegraphics[width=4.4cm,angle=90]{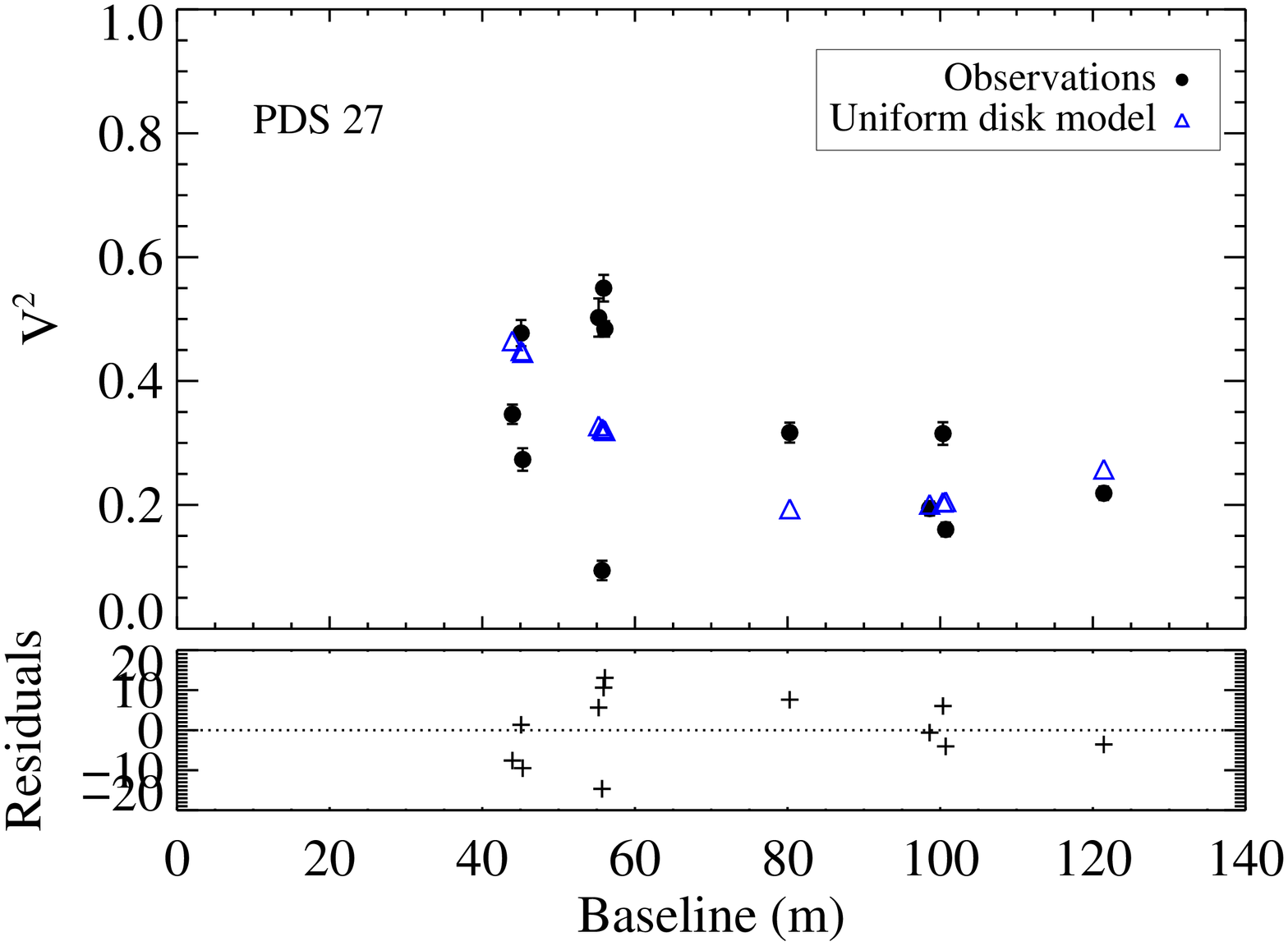} &     
                \includegraphics[width=4.4cm,angle=90]{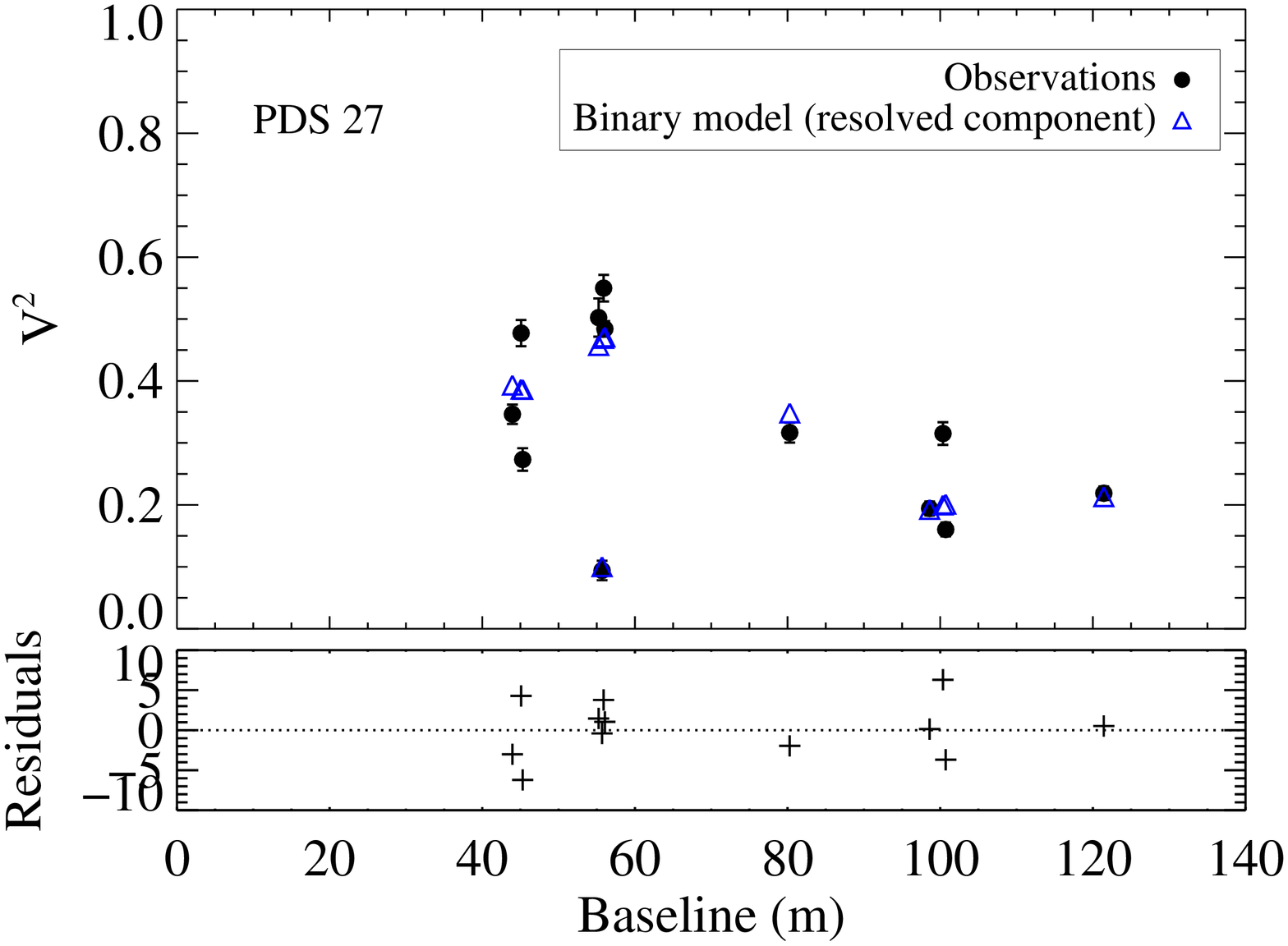} & \includegraphics[width=4.4cm,angle=90]{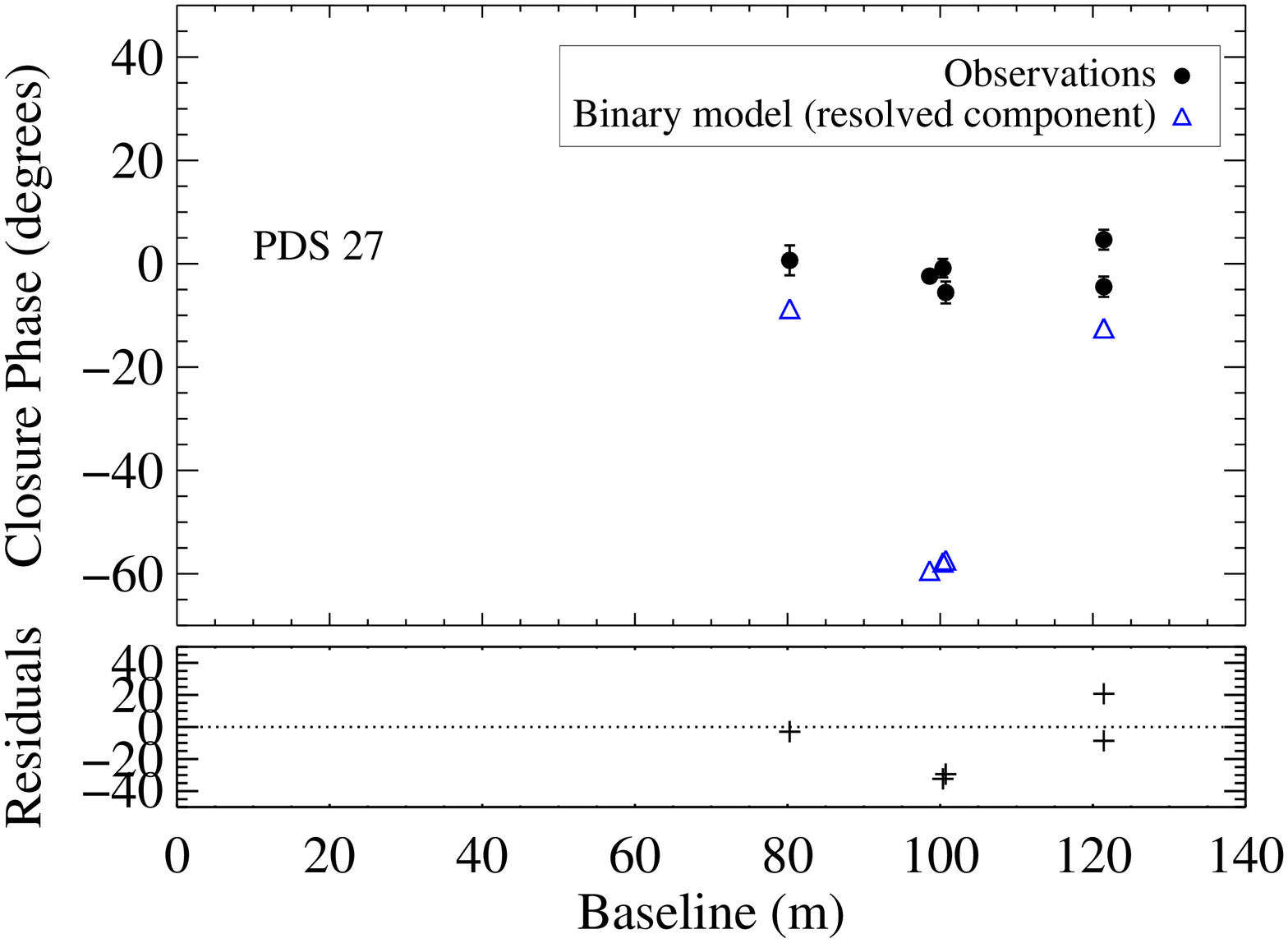} \\ 
\end{array}$
\end{center}
          \caption{Top: Three different models are compared with the observed visibility of PDS 37. The black circles with vertical error bars are the observed data, while the solid black triangles represent the best visibility fit. The best-fit models correspond to i) a uniform disk (left; ${\chi}^2$=140), ii) a point source binary (middle; ${\chi}^2$=220), and iii) a binary model with a resolved component (right; ${\chi}^2$=7.6). Bottom: Same as for PDS 37, but for PDS 27. Two different models are compared with the observed visibility of PDS 27, i) a uniform disk model (left; ${\chi}^2$=90) and ii) resolved binary model (right; ${\chi}^2$=16). For the best visibility fit we also show the predicted vs. observed closure phases at the maximum of their spatial frequency. The observed closure phases disagree with the models.}
   \label{27v}
       \end{figure*}

To begin with, we first considered a uniform disk model for a single object. The flux ratio between the central object and the disk and the size of the disk were set as free parameters. The best fit is shown in the top panel of Figure~\ref{27v}, and it shows a poor agreement between the observed visibilities and the model (${\chi}^2$ $\sim$ 140). 

Second, we considered fitting the data with a point source binary model without a circumstellar environment. In this case the primary object was set at a fixed central position, while the position of the secondary and the flux ratio of the two objects were set as free parameters. The best fit is shown in the middle panel of Figure~\ref{27v} and shows an unsatisfactory agreement with the observed data (${\chi}^2$ $\sim$ 220). However, in this case, the general shape of the measured visibility is in reasonable agreement with the observed visibility because both show fluctuations. These two models give a good indication that PDS 37 is likely a binary system in which one or both objects are resolved. 

Finally, we considered a binary system with one resolved component by adding a disk structure surrounding one object. The added complication of this situation is the large number of variables. For simplicity, we considered two stars (point sources) with a disk, and we fixed the position of one object and the disk at the origin and allowed the position of the second object to be fitted by the algorithm. The flux ratio (star+disk / star) and the size of the disk were also set as free parameters. This will give two possible results depending on the position of the secondary object and the diameter of the disk. The two possible results are a binary system with a circumstellar disk around both components or a disk around one object and a point source companion. The result for a binary and a disk around one object is shown in the top right panel of Figure~\ref{27v}. The figure shows that the model is able to better fit the visibility with a ${\chi}^2$ $\sim$ of 7.6 than the previous two models. 

After we determined the best fit of the visibilities, we proceeded with our fitting process by including the observed closure phases. Figure~\ref{27v} shows that although the last model provides a good fit for the visibilities, it fails to predict the observed closure phases. We were able to fit both visibilities and closure phases of PDS 37 when we adopted a binary with a) one resolved ring component, b) a binary with two resolved components (disks) and some asymmetric flux, or c) a binary with two resolved components (disks). These were our best-fit models, and they are presented in Sec.~\ref{geo} as models a, b, and c. In model a, we chose to build a geometric model using a ring surrounding the primary instead of a disk. During the initial fitting process, i) the position of the secondary object, ii) the flux ratio (ring+primary/secondary), iii) the geometric characteristics of the ring (i.e. PA, size, flattened ratio) were all let as free parameters. In model b) i) the flux ratios (primary+disk/secondary+disk), ii) the position of the secondary (both disk and point source), and iii) the size of the two disks were set as free parameters. The difference between models b and c is that in model c the position of the secondary object was fixed to be in the center of the secondary disk. For these three more complex models, one group of parameters (i.e., i, ii, and iii) was fixed one at a time after a first estimate of the parameters was achieved, and the other two groups were fit as a part of the fine-tuning process following an alternating sequence.


\begin{figure*}[ht]
\begin{center}
                \includegraphics[width=7cm]{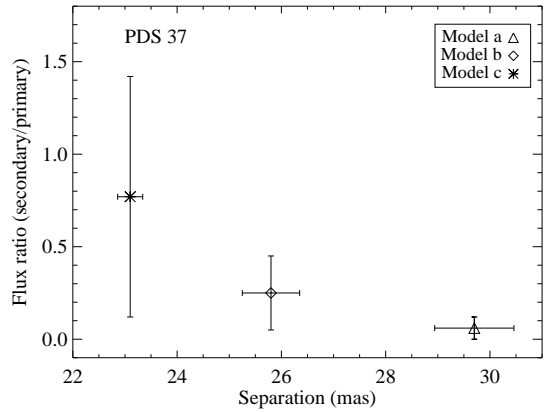}      
\end{center}
               \caption{Separations and flux ratios of PDS 37 when modelled as binary. The best fit binary models (a, b, c) are presented in Figure~\ref{models}. All binary models indicate separations of $\sim$22-28~mas which correspond to 42-54 au at the distance of the source. The flux ratios are characterized by a large uncertainty but we observe decreasing flux ratios with increasing separations.}
          \label{flux_dist_bin}
\end{figure*}

The best fit binary models for PDS 37 resulted in a range of flux ratios and separations. Figure~\ref{flux_dist_bin} presents the resulting flux ratios (secondary/primary) versus the separation of the two companions. We find that while the separation is reasonably constrained to be $\sim$ 22-28 mas (42-54 au) for PA of 80$^{\circ}$, 260$^{\circ}$ and 305$^{\circ}$, the flux ratios are highly uncertain and appear to decrease with increasing separation. 

\subsection{PDS 27}

The same procedure as for PDS 27 was followed as for PDS 37. The results of a disk model and a binary model with one resolved companion are presented in Figure~\ref{27v}; the left plot shows the disk model, and the right plot shows the binary model with one resolved object. The figure shows  a poor agreement between the observed visibility and the disk model. 

In contrast, the binary model provides a good fit to the observed visibility for the individual data. The position of the companion is found to be at $\sim$10~au separation at the distance of PDS 27. The best fit also indicates a disk of a diameter of 3.9$\pm$0.2 mas around the primary object. The flux ratio has a large uncertainty that is comparable to the value itself, but it appears that star/disk flux ratio is around 4\%, indicating that the resolved object is deeply embedded in its disk environment. The reduced ${\chi}^2$ map of the position of the companion shows several local minima with the minimum to be ${\chi}^2$ $\sim$ 16. This is a degenerate problem where several combinations of x and y lead to a good fit. Therefore, the available data are not sufficient to constrain all the parameters, and the observed degeneracy can be broken by obtaining more data. Figure~\ref{27v} also shows that similar to PDS 37, the model fails to predict the observed closure phases, and therefore a more detailed model is required. A summary of the fitting progression is presented in Table.~\ref{badfit}. 

The models that provide good fits for both visibilities and closure phases and are presented in Sec.~\ref{geo}. For PDS 27 the presence of a companion is not necessary when we fit the observables with a) one resolved ring, while b) a binary with two resolved components (disks) was the best solution when we treated a binary geometry. In model a, we considered a model where the nature of a resolved object was a ring instead of a disk. During the initial fitting process, i) the flux ratio (ring/star) and ii) the geometric characteristics of the ring (i.e., PA, size, and flattened ratio) were all set as free parameters. In model b), i) the flux ratios (primary+disk/secondary+disk), ii) the position of the secondary, and iii) the size of the two disks were set as free parameters. As in PDS 37, after the first estimate of the parameters, one group (i.e., i, ii, and iii) was fixed, and the other two groups were fit in an alternating sequence.

\begin{table*}[ht]
\caption{Parameters of the initial fit models tested for PDS 27 and PDS 37.}
\small
\centering
\setlength\tabcolsep{2pt}
\begin{tabular}{l c c c c c c c}
\hline\hline
{\bf{Model}} & flux weight 1 &  & flux weight disk & disk size & x2 & y2 & Red. ${\chi}^2$[DoF]$*$ \\ {\bf{Uniform disk}} & & & & mas & mas & mas &  \\
\hline
PDS 37 & $0.6^{+1.1}_{-0.6}$ &  & $0.4^{+0.8}_{-0.4}$ & 7.3$\pm$1.0 & 0 (fixed) & 0 (fixed) & 140 [9] \\
PDS 27 & $0.5^{+0.9}_{-0.5}$ &  & $0.5^{+0.9}_{-0.5}$ & 6.4$\pm$0.5 & 0 (fixed) & 0 (fixed) & 90 [9] \\
\hline
{\bf{Model}} & flux weight 1 & flux weight 2 &  &  & x2 & y2 & Red. ${\chi}^2$/DoF \\ {\bf{Binary (point sources)}} & & & &  & mas & mas &  \\
\hline
PDS 37 & $0.5^{+0.5}_{-0.5}$ & $0.5^{+0.5}_{-0.5}$ &  &  & 14.1$\pm$0.2 & 0.8$\pm$0.4 & 220 [8] \\
PDS 27 & $0.5^{+0.5}_{-0.5}$ & $0.5^{+0.5}_{-0.5}$ &  &  & 7.0$\pm$1.2 & -4.4$\pm$0.9 & 300 [8] \\
\hline
{\bf{Model}} & flux weight 1 & flux weight 2 & flux weight disk & disk size & x2 & y2 & Red. ${\chi}^2$/DoF \\ {\bf{Binary (resolved source)}} & & & & mas & mas & mas &  \\
\hline
PDS 37 & $0.535^{+0.578}_{-0.535}$ & $0.135^{+0.151}_{-0.135}$ & $0.33^{+0.36}_{-0.33}$ & 4.6$\pm$0.5 & -22.5$\pm$0.2 & y2=-3.8$\pm$0.3 & 7.6 [6] \\
PDS 27 & $0.48^{+0.56}_{-0.48}$ & $0.02^{+0.03}_{-0.02}$ & $0.5^{+0.58}_{-0.5}$ & 3.9$\pm$0.2 & 3.7$\pm$0.2 & -1.3$\pm$0.2 & 16 [6] \\
\hline

\end{tabular}

\tiny Notes: $*$ DoF stands for Degrees of Freedom. The reported reduced ${\chi}^2$ is the result of the fitting process when only visibilities are taken into account.                                                       
\label{badfit}  
\end{table*}

\section{Bandwidth smearing}

\label{smear}

In interferometric data of low spectral resolution (R < 50), the bandwidth smearing effects can influence the measured interferometric observables. In particular, the observed visibilities may show a tapering of the peak-to-peak amplitude \citep[see Fig. 3 by][]{Lachaume2013} mimicking the signature of resolved emission, even when the emission is in fact unresolved. For closure phases, the group-delay tracking errors are the main source of smearing bias and do not concern single spectral channel observations like those presented in this study. In binaries, a significant bandwidth smearing will likely affect the measured flux ratio, and it may mimic the spatial extension of the component(s). Therefore it is important to investigate its significance in our dataset and modeled geometries.

As presented in \citet{Davis2000}, bandwidth smearing is negligible under the condition $\beta$ $=$ $\pi$B$\theta$/R$\lambda$ $<<$ 1, where $\theta$ is the angular distance from the central object, B is the projected baseline, R is the spectral resolution, and $\lambda$ is the wavelength. We find that for our PIONIER dataset, the effects of bandwidth smearing can become significant at baselines B $>$ 90~m and binary separations greater than 8 mas. In particular, following the method described in \citet{Kraus2005}, we find that at the longest projected baseline B = 120m, the visibilities of PDS 27 ($\alpha$ = 12 mas) and PDS 37 ($\alpha$ = 23 mas) could be artificially lowered by a factor of 0.79 and 0.35, respectively. It appears that the observed modulation in the visibility profile at baselines $<$ 90~m is due to the binary nature of the sources, but the observed drop in the visibility at B $>$ 90~m is significantly affected by bandwidth smearing effects. Therefore our resulting flux ratios and the extension of the binary component(s) suffer from this additional uncertainty. Fitting only the subset of the data where the bandwidth smearing should be negligible (e.g., UT1-UT2 and UT2-UT3) would be a good way to confirm the resulting parameters of our models. This would reduce the total amount of our data by half, resulting in 0 degrees of freedom in our models, which would make it impossible to proceed.

In conclusion, the binary nature of the objects, which is the main finding of this paper, is not affected by the bandwidth smearing. In future work, where the focus will be on the accurate determination of the flux ratios and the extension of the binary components by obtaining more interferometric data, the smearing effects will be treated more carefully \citep[e.g.,][]{Lachaume2013}.

\end{appendix}
\end{document}